\documentclass[aps,twocolumn,showpacs,superscriptaddress,floatfix]{revtex4}
\usepackage{graphicx,amsmath,amssymb,times}
\usepackage{txfonts}

\begin{document}

  \title{Conventional character of the BCS-BEC cross-over in ultra-cold
    gases of $^{40}$K
    %BCS-BEC cross-over in ultra-cold gases of $^{40}$K is conventional
  }

  \author{Marzena H. Szyma{\'n}ska}
  \affiliation{Cavendish Laboratory, Department of Physics,
    University of Cambridge, Madingley Road, Cambridge CB3 OHE, UK}
  \author{Krzysztof G{\'o}ral}
  \affiliation{Clarendon Laboratory, Department of Physics,
    University of Oxford, Parks Road, Oxford, OX1 3PU, UK}
  \author{Thorsten K{\"o}hler}
  \affiliation{Clarendon Laboratory, Department of Physics,
    University of Oxford, Parks Road, Oxford, OX1 3PU, UK}
  \author{Keith Burnett}
  \affiliation{Clarendon Laboratory, Department of Physics,
    University of Oxford, Parks Road, Oxford, OX1 3PU, UK}

\begin{abstract}

  We use the standard fermionic and boson-fermion Hamiltonians to study
  the BCS-BEC cross-over near the 202 G resonance in a two-component
  mixture of fermionic $^{40}$K atoms employed in the experiment of
  C.A. Regal {\it et al.} [Phys. Rev. Lett. \textbf{92}, 040403 (2004)].
  Our mean-field analysis of many-body equilibrium quantities 
  shows virtually no differences between the predictions of the two
  approaches, provided they are both implemented in a manner that properly 
  includes the effect of the highest excited bound state of the background 
  scattering potential, rather than just the magnetic-field dependence of the 
  scattering length. 
  %To this end, we show how the parameters of the
  %two Hamiltonians can be unambiguously determined from the microscopic 
  %two-body interactions. In contrast to previous predictions, we demonstrate 
  %that the population of the molecular
  %field of the boson-fermion Hamiltonian reaches a maximum of only 8\%
  %at 7.2 G below the zero-energy resonance, while it remains smaller
  %than 0.6\% above it. 
  Consequently, we rule out the macroscopic occupation of the molecular field 
  as a mechanism behind the fermionic pair condensation and show that  
  the BCS-BEC cross-over in ultra-cold $^{40}$K  gases can be analysed and 
  understood on the same basis as in the conventional systems of solid state 
  physics.

\end{abstract}
\date{\today}
\maketitle

\section{Introduction}
\label{sec:intro}
Dilute vapours of fermionic atoms have recently attracted considerable
attention as systems for studying
%possible candidates for environments to study
the cross-over between Bardeen-Cooper-Schrieffer (BCS) pairing and
Bose-Einstein condensation (BEC) of self-bound pairs of particles
\cite{Keldysh,Eagles,Leggett,Noz-SchR,Rand}. The experimental techniques
underlying these studies all take advantage of the unique
possibility of controlling the inter-atomic interactions via
magnetically tunable Fesh\-bach resonances. Observations of both
molecular BEC \cite{Greiner03,Grimm-mol,Ket-mol} and condensation of
unbound fermionic pairs \cite{Regal04-2,Zwierlein04} in gases
containing incoherent mixtures of two different spin components have
recently been reported.  The properties of both phases have been
studied in several recent experiments
\cite{Bartenstein04,Bourdel04,Kinast04,Chin04,Greiner04,Zwierlein05}.

The principal question of whether the BCS-BEC cross-over in
ultra-cold atomic gases can be analysed and understood on the same
basis as in the traditional systems of solid state physics
\cite{Keldysh,Eagles,Leggett,Noz-SchR,Rand} is, however, a matter of 
continuing controversy.  This controversy dates back to the first 
predictions of a superfluid phase transition in dilute Fermi gases using 
Fesh\-bach resonances \cite{Holland,Timm}, suggesting a new type of pair
condensation based on a macroscopic occupation of the Fesh\-bach
resonance level. The concept underlying this idea of resonance
superfluidity \cite{Holland,Timm} was motivated by the use of an
effective boson-fermion Hamiltonian \cite{Friedberg89,Ranninger85},
which treats pairs of atoms in the resonance state configuration in
terms of single structureless Bose particles, often referred to as
``molecules''. The studies of Refs.~\cite{KGB03,TKTGPSJKB03} have
clearly shown, however, that the experimentally relevant highest
excited vibrational multi-channel molecular bound state (the
Fesh\-bach molecule) is, in general, significantly different from the
Fesh\-bach resonance level (belonging to the closed scattering
channel) and requires an explicit description in
terms of a composite two-body system.

In this paper we demonstrate that the thermal equilibrium physical quantities, 
relevant to recent experiments on the BCS-BEC cross-over in cold gases 
using broad Fesh\-bach resonances of $^{40}$K and $^6$Li, can all be described 
by the usual fermionic Hamiltonian for a gas with two spin components and 
separable binary interactions
\cite{Leggett,Noz-SchR,Schrieffer}. Consequently, we rule out the
macroscopic occupation of the Fesh\-bach resonance level as a mechanism
behind the superfluid pairing in these systems. To this end, 
we compare predictions obtained from both the standard fermionic and 
boson-fermion Hamiltonians adjusted in such a way that they both recover the 
low energy binary collision physics over a wide range of magnetic field 
strengths, beyond the regime of universality. These adjustments are performed 
on the basis of a two-channel description of the resonance enhanced binary 
scattering involving the entrance channel of the spatially separated atoms and 
a closed channel strongly coupled to it via the Fesh\-bach resonance level. 
Our approach \cite{GKGTJ03} depends on five measurable parameters of a 
Fesh\-bach resonance and is applicable to both narrow 
(closed channel dominated) and broad (entrance channel dominated) resonances.

Throughout this paper we shall discuss a balanced mixture of $^{40}$K 
atoms prepared in the $(f=9/2,m_f=-9/2)$ and $(f=9/2,m_f=-7/2)$ Zeeman states 
(cf., e.g., Ref.~\cite{Regal04-2}). We consider magnetic field strengths 
in the vicinity of the 202 G (1 G $=10^{-4}$ T) zero-energy resonance 
(singularity of the scattering length). This $^{40}$K Fesh\-bach resonance is 
particularly well suited to demonstrate the wide range of applicability of our 
approach, as its width is in between those of the experimentally relevant 
extremely narrow (closed channel dominated) and broad (entrance channel 
dominated) resonances of $^{6}$Li at about 543 \cite{Strecker03} and
830 G \cite{Grimm-mol,Ket-mol,Bourdel04,Kinast04}, respectively. While 
the general differences between closed and entrance channel dominated 
Fesh\-bach resonances were discussed in Ref.~\cite{KGG04}, the very different 
universal regimes of magnetic field strengths of $^6$Li resonances and their 
potential relevance to the BCS-BEC cross-over were the subject of the recent 
studies of Ref.~\cite{Strinati}. 

The paper is organised as follows: In Section \ref{sec:two-body} we
describe the two-channel approach to the resonance enhanced low energy
binary scattering observables and introduce the five relevant physical
parameters for the 202 G zero-energy resonance of $^{40}$K. These
two-body considerations reveal that the admixture of the Fesh\-bach
resonance level to the experimentally relevant bound Fesh\-bach
molecule never exceeds 8\% over the entire range of magnetic fields
relevant to the experiment of Ref.~\cite{Regal04-2}. Its maximum value
is reached at about 7.2 G below the position of $B_0=202.1$ G
\cite{Regal04-2} of the zero-energy resonance. These results are in
agreement with exact coupled channels calculations \cite{Julienne}
from which the two-channel approach was originally derived in
Ref.~\cite{GKGTJ03}. Our analysis reveals that a proper description of
the admixture of the resonance level to the multi-channel bound and
scattering states requires a two-channel approach to explicitly
account for the entrance channel background scattering potential
including its highest excited vibrational state. Given the small
admixture of the resonance level to the Fesh\-bach molecule, we then
determine an accurate separable single channel potential, which is
suitable for the standard fermionic many-body Hamiltonian. This
potential is adjusted in such a way that it accurately describes not
only the magnetic field dependence of the scattering length but also
the binding energy of the Fesh\-bach molecule.

In Section \ref{sec:model} we discuss the implications of the nature of
the resonance enhanced binary collision physics for the potentials
and parameters of both the standard fermionic and boson-fermion
Hamiltonians. The procedure of adjusting these effective Hamiltonians 
involves the identification of the
two-body resonance level with the structureless Bose particles of the
boson-fermion approach. Its background scattering potential and
inter-channel coupling parameters are then determined in such a way that
they exactly reproduce the predictions of the two-channel approach of
Ref.~\cite{GKGTJ03} when applied to a pair of atoms.

Section \ref{sec:therm} briefly summarises the mean-field approach we
have used to determine the thermodynamic quantities for both the
standard fermionic and the boson-fermion Hamiltonians. In Section
\ref{sec:Results} we then apply this mean-field approach to the
BCS-BEC cross-over in a cold gas of $^{40}$K with a particular
emphasis on the experimental conditions of Ref.~\cite{Regal04-2}. We
show that, as expected from our two-body analysis, the standard
fermionic and boson-fermion Hamiltonians lead to virtually the same
predictions about the equilibrium physics. In particular, our analysis
supports the picture of a smooth cross-over of the pair size in the
entrance channel characterised by features similar to those of
traditional superconductivity \cite{Keldysh,Eagles,Leggett,Noz-SchR,Rand} and
exciton and polariton \cite{Keldysh,Littlewood04,Eastham01,Keeling04}
condensates. We therefore conclude that the occupation of the Fesh\-bach
resonance level is irrelevant to the nature
of fermionic superfluidity in a cold $^{40}$K gas, in contrast to the findings
of Refs.~\cite{Stoof,Mackie05}. We predict the position of the zero of the 
chemical potential (cross-over point) at 0.32 G below the zero-energy
resonance. We have, furthermore, analysed the experimental pairwise
projection technique of Ref.~\cite{Regal04-2} by calculating the
overlap between the fermionic pairs produced on the high field side of
the $^{40}$K resonance and the bound Fesh\-bach molecule on its low
field side. Our results indicate that fermionic pair condensation
phenomena should become invisible once the gas is prepared at magnetic
field strengths further than about 0.5 G above the resonance position.
Finally, Section \ref{sec:concl} summarises our main conclusions.

\section{Magnetically tunable inter-atomic interactions}
\label{sec:two-body}

\begin{figure}[!htbp]
  \begin{center}
    \includegraphics[width=\columnwidth,clip]{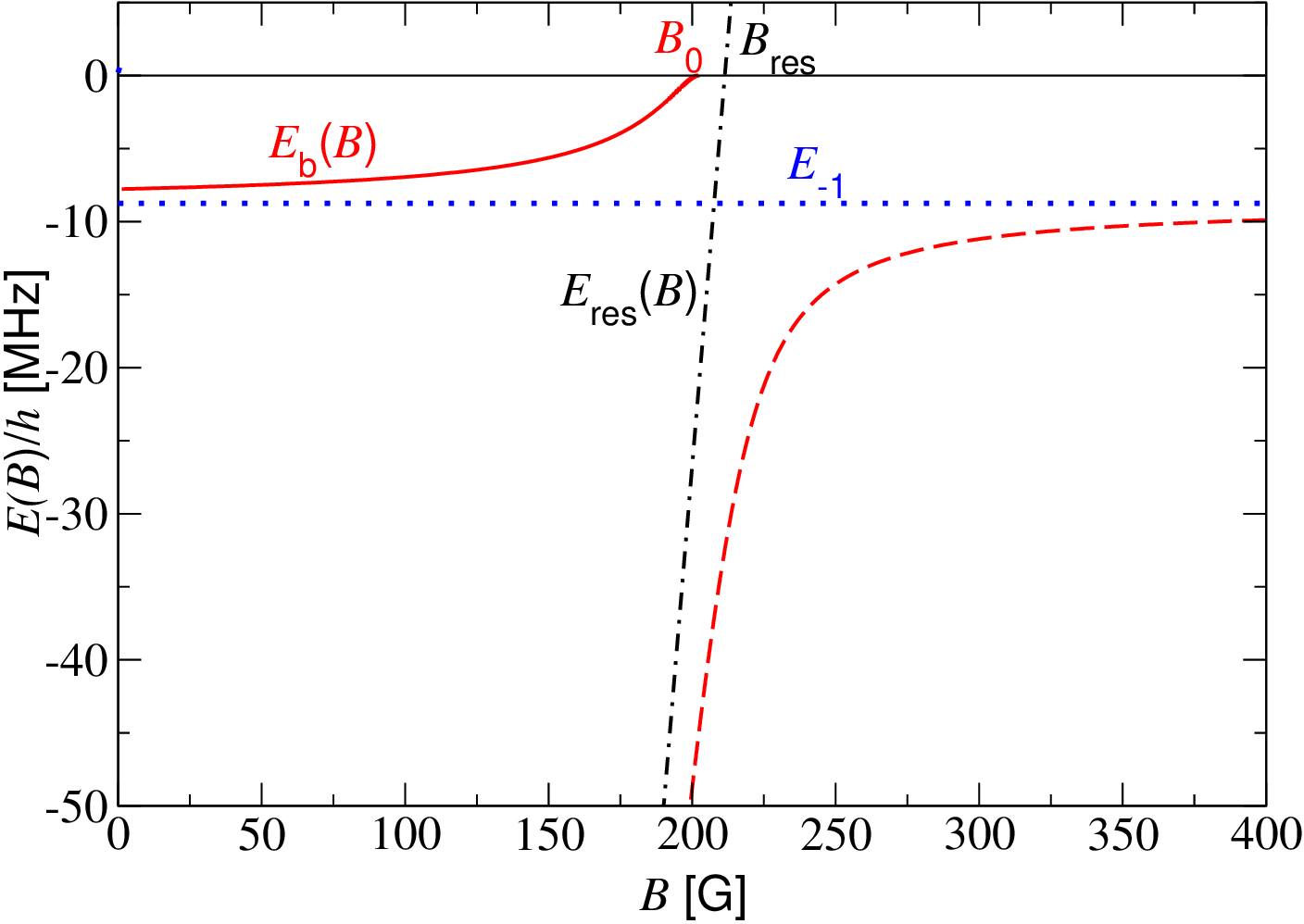}
    \caption{(Color online) Binding energies of the highest excited
      multi-channel vibrational molecular bound states (solid and dashed 
      curves), associated with a pair of unlike $^{40}$K atoms in the
      asymptotic $(f=9/2,m_f=-9/2)$ and $(f=9/2,m_f=-7/2)$ Zeeman states,
      versus the magnetic field strength $B$. The Fesh\-bach molecular state 
      $\phi_\mathrm{b}(B)$ and its energy $E_\mathrm{b}(B)$ emerge 
      at the position of the zero-energy resonance $B_0$. The linearly varying 
      energy $E_\mathrm{res}(B)$ of the Fesh\-bach resonance level 
      $\phi_\mathrm{res}$ and the energy $E_{-1}$ of the highest excited 
      vibrational state of the background scattering potential are indicated 
      by the dot-dashed and dotted lines, respectively. We note that the 
      measurable resonance position $B_0$ is significantly shifted with
      respect to the magnetic field strength $B_\mathrm{res}$ at which
      the closed-channel resonance state energy crosses the dissociation 
      threshold of the 
      entrance channel. Here and throughout this paper, we have chosen the 
      zero of the energy, at each magnetic field strength $B$, as the sum of 
      the Zeeman energies of a pair of unlike fermions.}
    \label{fig:E-1}
  \end{center}
\end{figure}

At the low collision energies characteristic of cold dilute gases the binary 
scattering physics can be described by a single parameter of the inter-atomic 
potential, the $s$-wave scattering length $a$. The experimental technique 
of Fesh\-bach resonances takes advantage of the Zeeman effect in the 
atomic energy levels to widely tune the scattering length using magnetic 
fields. In the experiments of Ref.~\cite{Regal04-2} the gas was prepared as a 
balanced, incoherent mixture of the lowest energetic $(f=9/2,m_f=-9/2)$ 
and $(f=9/2,m_f=-7/2)$ Zeeman levels at magnetic field strengths on the order 
of 200 G. Only pairs of unlike fermions in different Zeeman levels can 
interact via $s$ waves which dominate cold collisions (in the absence of 
resonant phenomena associated with higher partial waves). Throughout this 
paper we shall denote the asymptotic $s$-wave binary scattering channel of 
a pair of these unlike $^{40}$K atoms as the entrance channel, while its 
associated interaction potential $V_\mathrm{bg}$ will be referred to as the 
background scattering potential. As the $m_f$ degeneracy of the Zeeman 
levels is removed by a homogeneous magnetic field of strength $B$, the 
potentials associated with the different scattering channels can be shifted 
with respect to each other. The typically weak inter-channel coupling becomes 
significantly enhanced when the magnetic field dependent energy 
$E_\mathrm{res}(B)$ (see Fig.~\ref{fig:E-1}) of a closed channel vibrational 
state $\phi_\mathrm{res}$ (the Fesh\-bach resonance level) is tuned in the 
vicinity of the dissociation threshold of the entrance channel. Since this 
threshold coincides with the zero of collision energy 
(zero of energy in Fig.~\ref{fig:E-1}), 
its virtual degeneracy with $E_\mathrm{res}(B)$ leads to a resonance 
enhancement of the cold inter-atomic collisions in terms of a zero-energy 
resonance in the entrance channel, i.e.~a singularity of the scattering length
at the magnetic field strength $B_0$ described by the formula:
\begin{align}
  a(B)=a_\mathrm{bg}\left(1-\frac{\Delta B}{B-B_0}\right). 
  \label{aofB}
\end{align}
The parameters $a_\mathrm{bg}$ and $\Delta B$ are usually referred to as 
the background scattering length and the resonance width, respectively.

The magnetic tuning of the energy of the Fesh\-bach resonance level distorts 
not only the scattering continuum but also the bound multi-channel molecular 
energy levels. As illustrated in Fig.~\ref{fig:E-1}, the position $B_0=202.1$ 
G of the zero-energy resonance, i.e.~the magnetic field strength at which 
the scattering length has its singularity, coincides with the zero of the 
binding energy $E_\mathrm{b}(B)$ of the highest excited multi-channel 
vibrational state. This Fesh\-bach molecular state $\phi_\mathrm{b}(B)$ 
persists on the low field side of the 202 G zero-energy resonance of 
$^{40}$K where $a(B)$ is positive, while it ceases to exist on the high field 
side of negative scattering lengths. We note that the measurable resonance 
position $B_0$ is shifted with respect to the magnetic field strength 
$B_\mathrm{res}$ (see Fig.~\ref{fig:E-1}) at which the energy 
$E_\mathrm{res}(B)$ crosses the dissociation threshold of the entrance 
channel. In fact, the multi-channel bound state $\phi_\mathrm{b}(B)$ and the 
meta-stable Fesh\-bach resonance level are, in general, rather distinct 
objects, in particular, when the inter-channel coupling is strong.
 
\subsection{Two-channel approach}

To accurately describe the resonance enhancement of the 
binary collisions as well as the properties of the Fesh\-bach molecule of a 
pair of unlike $^{40}$K fermions, we employ the approach of 
Ref.~\cite{GKGTJ03}, which explicitly includes the entrance channel and its
strong coupling to a closed channel via the Fesh\-bach resonance level. 
According to the effective treatment of Ref.~\cite{GKGTJ03}, the general 
form of a two-channel Hamiltonian matrix
\begin{align}
  H_\mathrm{2B-2ch}=
  \left(
  \begin{array}{cc}
    -\frac{\hbar^2}{m}\nabla^2+V_\mathrm{bg} & W\\
    W & -\frac{\hbar^2}{m}\nabla^2+V_\mathrm{cl}(B)
  \end{array}
  \right) 
  \label{H2B2channel}
\end{align}
depends on five physical parameters in addition to the atomic mass $m$: 
The background scattering potential $V_\mathrm{bg}$ of the entrance 
channel is characterised by the energy $E_{-1}$ of its highest excited 
vibrational state and the background scattering length $a_\mathrm{bg}$
\cite{Gao98}, while the closed channel potential $V_\mathrm{cl}(B)$ can be 
effectively described by the constant difference $dE_\mathrm{res}/dB$ between 
the magnetic moments of the Fesh\-bach resonance level and a pair of 
asymptotically separated $^{40}$K fermions in the $(f=9/2,m_f=-9/2)$
and $(f=9/2,m_f=-7/2)$ Zeeman states. The inter-channel coupling is 
accounted for by the off-diagonal potential $W$ whose strength and range
parameter are determined by the resonance width $\Delta B$ and the
shift $B_0-B_\mathrm{res}$ (see Fig.~\ref{fig:E-1}).

At the low relative momenta and associated large de Broglie wavelengths 
characteristic for cold gases, the details of the microscopic binary 
interaction potentials are not resolved. Reference \cite{GKGTJ03}, therefore, 
suggests a convenient effective potential matrix that recovers all the 
relevant low energy physical observables of the exact microscopic binary 
interaction. 

\subsubsection{Adjustment of the background scattering potential}
Following the treatment of Ref.~\cite{GKGTJ03}, we use the following separable 
form of the background scattering potential:
\begin{equation}
  V_\mathrm{bg}=|\chi_\mathrm{bg}\rangle \xi_\mathrm{bg}
  \langle\chi_\mathrm{bg}| \, .
  \label{separablepotentialgeneral}
\end{equation}
For the form factor $|\chi_\mathrm{bg}\rangle$ we choose the Gaussian 
{\em ansatz}:
\begin{equation}
  \langle\mathbf{p}|\chi_\mathrm{bg}\rangle=\chi_\mathrm{bg}(p)=
  \frac{\exp\left(-\frac{p^2\sigma_\mathrm{bg}^2}{2\hbar^2}\right)}
       {(2\pi\hbar)^{3/2}} \, .
       \label{separablepotentialGaussian}
\end{equation}
Here $|\mathbf{p}\rangle$ is a plane wave momentum state, normalised
as $\exp(i\mathbf{p}\cdot\mathbf{r}/\hbar)/(2\pi\hbar)^{3/2}$.
The amplitude $\xi_\mathrm{bg}$ and the range parameter 
$\sigma_\mathrm{bg}$ can be adjusted in such a way that 
$V_\mathrm{bg}$ exactly recovers $a_\mathrm{bg}$ and $E_{-1}$ as follows:
The first condition for the adjustments is given by the relationship between 
the background scattering length and the zero-energy $T$ matrix associated 
with the background scattering potential. The exact analytic expression for 
this $T$ matrix given in Ref.~\cite{GKGTJ03} yields:
\begin{align}
  a_\mathrm{bg}=\frac{\frac{m}{4\pi\hbar^2}\xi_\mathrm{bg}}
  {1+\frac{m}{4\pi^{3/2}\hbar^2\sigma_\mathrm{bg}}\xi_\mathrm{bg}}.
  \label{condabg}
\end{align}
The second condition is equivalent to the stationary Schr\"o\-dinger 
equation for a bound state of the separable potential and reads 
\cite{GKGTJ03}:
\begin{align}
  \frac{m}{4\pi^{3/2}\hbar^2\sigma_\mathrm{bg}}\xi_\mathrm{bg}
  \left[\sqrt{\pi}xe^{x^2}\mathrm{erfc}(x)-1\right]=1.
  \label{condEminusone}
\end{align}
Here $\mathrm{erfc}(x)=\frac{2}{\sqrt{\pi}}\int_x^\infty e^{-u^2} du$ is the 
complementary error function (sometimes referred to as the Hechenblaikner 
function) with the argument $x=\sqrt{m|E_{-1}|}\sigma_\mathrm{bg}/\hbar$. We 
note that the positivity of the background scattering length 
$a_\mathrm{bg}=174\, a_\mathrm{Bohr}$ \cite{Loftus02} 
($a_\mathrm{Bohr}=0.052918$ nm is the Bohr radius) guarantees the existence of 
a single bound state of the separable potential $V_\mathrm{bg}$ for two unlike 
$^{40}$K fermions. Equations (\ref{condabg}) and (\ref{condEminusone}) then
simultaneously determine the parameters $\sigma_\mathrm{bg}$ and 
$\xi_\mathrm{bg}$.

\subsubsection{Single-resonance approximation in the closed channel}
In most of the experimentally relevant cases the entrance channel is strongly 
coupled only to a single closed channel vibrational state, i.e.~the Fesh\-bach 
resonance level $\phi_\mathrm{res}$. It is therefore usually
sufficient to restrict the spatial configuration of a closed channel atom pair 
to $\phi_\mathrm{res}(r)$, where $r$ is the inter-atomic separation. To this 
end Ref.~\cite{GKGTJ03} introduces the single-resonance approximation
(referred to as the pole approximation in Ref.~\cite{GKGTJ03}) which consists 
in replacing the diagonal closed channel part of the general Hamiltonian 
matrix (\ref{H2B2channel}) by the one dimensional projection onto the
resonance level, i.e.
\begin{align}
  -\hbar^2\nabla^2/m+V_\mathrm{cl}(B)\to
  |\phi_\mathrm{res}\rangle E_\mathrm{res}(B)\langle\phi_\mathrm{res}|.
\end{align}
The resonance energy varies virtually linearly with the magnetic field 
strength $B$ and can therefore be described by the first term of the 
power series expansion about its zero at $B_\mathrm{res}$, which reads:
\begin{equation}
  E_\mathrm{res}(B)=\frac{dE_\mathrm{res}}{dB}(B-B_\mathrm{res})\, .
  \label{Eres}
\end{equation}
Here $dE_\mathrm{res}/dB$ is the magnetic moment of the resonance level 
with respect to the sum of magnetic moments of a pair of asymptotically
separated unlike fermions. It can be obtained either from measurements of 
the binding energies of the Fesh\-bach molecule away from the zero-energy 
resonance or from exact coupled channels calculations using microscopic 
interactions.

\subsubsection{Inter-channel coupling}
In accordance with the preceding single-resonance approximation to the 
closed channel Hamiltonian the inter-channel coupling is determined by 
the product $W|\phi_\mathrm{res}\rangle$ \cite{GKGTJ03}. We represent 
this product in terms of an amplitude $\zeta$ and a wave function 
$|\chi\rangle$, i.e.
\begin{equation}
  W|\phi_\mathrm{res}\rangle=|\chi\rangle\zeta.
  \label{coupling}
\end{equation}
The wave function $|\chi\rangle$ accounts for the spatial variation of the 
inter-channel coupling in terms of a range parameter $\sigma$. We choose a 
Gaussian {\em ansatz}\, which, in the convenient momentum space 
representation, is given by:
\begin{equation}
  \langle\mathbf{p}|\chi\rangle=\chi(p)=
  \frac{\exp\left(-\frac{p^2\sigma^2}{2\hbar^2}\right)}
       {(2\pi\hbar)^{3/2}}\, .
       \label{couplingGaussian}
\end{equation}

Given that the range parameter $\sigma_\mathrm{bg}$ of the 
background scattering potential has been determined using the procedure 
from above, the parameters $\zeta$ and $\sigma$ can be adjusted in such 
a way that they exactly recover the width $\Delta B$ of the zero-energy 
resonance as well as its shift $B_0-B_\mathrm{res}$ (see Fig.~\ref{fig:E-1}). 
Following the procedure of Ref.~\cite{GKGTJ03}, the two conditions for the 
simultaneous adjustment of $\zeta$ and $\sigma$ are given by the exact 
expressions 
\begin{align}
  \Delta B=\frac{m\zeta^2}{4\pi\hbar^2a_\mathrm{bg}(dE_\mathrm{res}/dB)}
  \left(
  1-\frac{a_\mathrm{bg}}{\sqrt{\pi}\overline{\sigma}}
  \right)^2
  \label{deltaBofzetasigma}
\end{align}
for the resonance width, and 
\begin{align}
  B_0-B_\mathrm{res}=(\Delta B)\frac{a_\mathrm{bg}}{\sqrt{\pi}\sigma}
  \frac{1-\frac{a_\mathrm{bg}}{\sqrt{\pi}\sigma}
    \left(\frac{\sigma}{\overline{\sigma}}\right)^2}
       {\left(
	 1-\frac{a_\mathrm{bg}}{\sqrt{\pi}\sigma}
	 \frac{\sigma}{\overline{\sigma}}
	 \right)^2}
       \label{shiftozetasigma}
\end{align}
for the associated shift. Here we have introduced the mean range parameter:
\begin{equation}
  \overline{\sigma}=
  \sqrt{\frac{1}{2}
    \left(
    \sigma^2+\sigma_\mathrm{bg}^2
    \right)}\, .
\end{equation}

While $\Delta B$ is usually known from measurements of the magnetic field 
dependence of the scattering length $a(B)$ (cf.~Ref.~\cite{Greiner03}), the 
precise value of $B_0-B_\mathrm{res}$ can not be easily deduced from 
experimental data. Reference \cite{GKGTJ03} therefore suggests an analytic 
treatment, based on ideas of multi-channel quantum defect theory 
\cite{Julienne89}, to determine the resonance shift. This treatment relates
$B_0-B_\mathrm{res}$ to the van der Waals dispersion coefficient $C_6$ in 
terms of the mean scattering length \cite{Gribakin93}:
\begin{align}
  \overline{a}=\frac{1}{\sqrt{2}}
  \frac{\Gamma(3/4)}{\Gamma(5/4)}
  \frac{1}{2}\left(\frac{mC_6}{\hbar^2}\right)^{1/4}.
  \label{abar}
\end{align}
Here $\Gamma$ is the gamma function.
The van der Waals dispersion coefficient characterises the (non-retarded) 
asymptotic behaviour $V_\mathrm{bg}(r)\underset{r\to\infty}{\sim}-C_6/r^6$ of 
the exact microscopic background scattering potential at large inter-atomic 
distances $r$ and can be deduced from experimental observations
\cite{Derevianko99}. Following the ideas of Ref.~\cite{Julienne89}, the 
magnitude of $B_0-B_\mathrm{res}$ is accurately determined by the formula:
\begin{equation}
  B_0-B_\mathrm{res}=(\Delta B)\frac{\frac{a_\mathrm{bg}}{\overline{a}}
    \left(1-\frac{a_\mathrm{bg}}{\overline{a}}\right)}
  {1+\left(1-\frac{a_\mathrm{bg}}{\overline{a}}\right)^2}.
  \label{magicformula}
\end{equation}
The procedure for the determination of the inter-channel coupling 
parameters $\zeta$ and $\sigma$ thus consists in calculating the 
resonance shift using Eq.~(\ref{magicformula}) and then simultaneously 
solving Eqs.~(\ref{deltaBofzetasigma}) and (\ref{shiftozetasigma}).

\subsubsection{Molecular bound states}
The two-channel Hamiltonian of our approach of Ref.~\cite{GKGTJ03} can support 
two bound states. Their wave functions associated with the entrance- and 
closed-channel components have the following general form:
\begin{equation}
  \left(
  \begin{array}{c}
    \phi_\mathrm{b}^\mathrm{bg}\\
    \phi_\mathrm{b}^\mathrm{cl}
  \end{array}
  \right)=\frac{1}{\mathcal{N}_\mathrm{b}}
  \left(
  \begin{array}{c}
    G_\mathrm{bg}(E_\mathrm{b})W\phi_\mathrm{res}\\
    \phi_\mathrm{res}
  \end{array}
  \right)\,.
  \label{phib}
\end{equation}
Here $G_\mathrm{bg}(E_\mathrm{b})$ is the energy-dependent Green's function 
associated with the entrance channel, whose explicit expression reads
\begin{equation}
  \label{Gbg}
  G_\mathrm{bg}(E_\mathrm{b})=
  \left[
    E_\mathrm{b}-\left(-\frac{\hbar^2}{m}\nabla^2+V_\mathrm{bg}\right)
    \right]^{-1}\,,
\end{equation} 
while the normalisation constant $\mathcal{N}_\mathrm{b}$ is given by the 
formula
\begin{equation}
  \mathcal{N}_\mathrm{b}=\sqrt{1+
    \langle\phi_\mathrm{res}|W
    \left[G_\mathrm{bg}(E_\mathrm{b})\right]^2W|\phi_\mathrm{res}\rangle}\,.
  \label{twochannelnormalisation}
\end{equation}
The energies $E_\mathrm{b}$ associated with the two-channel bound states in 
the single resonance approximation are determined by the following condition: 
\begin{equation}
  E_\mathrm{b}=E_\mathrm{res}(B)+
  \langle\phi_\mathrm{res}|WG_\mathrm{bg}(E_\mathrm{b})W|
  \phi_\mathrm{res}\rangle\,.
  \label{determinationEb}
\end{equation}
The energies of the Fesh\-bach molecular state and of the second 
more tightly bound two-channel state of $^{40}$K are indicated in 
Fig.~\ref{fig:E-1} by the solid and dashed curves, respectively.

\subsubsection{Parameters for a pair of unlike $^{40}$K fermions}
For the experiment of Ref.~\cite{Regal04-2}, which uses a balanced incoherent 
mixture of $^{40}$K atoms in the $(f=9/2,m_f=-9/2)$ and $(f=9/2,m_f=-7/2)$ 
Zeeman states at magnetic field strengths in the vicinity of the 202 G 
resonance, the physical parameters are given by: 
$a_\mathrm{bg}=174\, a_\mathrm{Bohr}$
\cite{Loftus02}, $E_{-1}=-h\times8.75$ MHz \cite{Julienne}, 
$\Delta B=7.8$ G \cite{Greiner03}, $B_0-B_\mathrm{res}=-9.274$ G, 
and $dE_\mathrm{res}/dB=1.679\, \mu_\mathrm{Bohr}$ 
($\mu_\mathrm{Bohr}=9.27400899\times 10^{-28}$ J/G is the Bohr 
magneton). We have obtained the resonance shift $B_0-B_\mathrm{res}$ 
from Eq.~(\ref{magicformula}) using $C_6=3897$ a.u. 
\cite{Derevianko99} (1\,a.u.=$0.095734\times10^{-78}$J m$^6$)
and the resonance slope $dE_\mathrm{res}/dB$ has been extracted from 
exact coupled channels calculations of the binding energy of the Fesh\-bach 
molecule \cite{Julienne}. From these physical quantities we have determined 
the following parameters of the two-channel Hamiltonian: 
$m\xi_\mathrm{bg}/(4\pi\hbar^2)=-244.852 \, a_\mathrm{Bohr}$,
$\sigma_\mathrm{bg}=57.387 \, a_\mathrm{Bohr}$, $\sigma=33.123 \,
a_\mathrm{Bohr}$, and $m\zeta^2/(4\pi\hbar^2\sigma)=h\times80.272$
MHz. Figure \ref{fig:E-1} shows the binding energies of the multi-channel
vibrational states predicted by the two-channel Hamiltonian of
Eq.~(\ref{H2B2channel}) using the parameters listed above. We note that 
the two-channel approach also exactly recovers the magnetic field dependence
of the scattering length $a(B)$ described by Eq.~(\ref{aofB}).

\subsection{Single-channel approach}
As the magnetic field strength approaches the position $B_0$ of the
zero-energy resonance from the positive scattering length side, the 
admixture of the resonance level $\phi_\mathrm{res}$ to the physically 
relevant Fesh\-bach molecular state $\phi_\mathrm{b}(B)$ vanishes 
\cite{TKTGPSJKB03,KGG04}. In this universal regime of magnetic field 
strengths it is always possible to introduce an effective single-channel 
Hamiltonian \cite{KGB03,GKGTJ03,KGG04}, which accurately recovers the 
resonance enhancement of the scattering as well as the universal binding 
energy 
\begin{align}
  E_\mathrm{b}\underset{a\to\,+\infty}{\sim}-\frac{\hbar^{2}}{m a^2} 
  \label{Ebuniversal}
\end{align}
of the Fesh\-bach molecule. The studies presented in Refs.~\cite{KGB03,KGG04} 
have shown, however, that the range of validity of such a single-channel 
approach can be significantly extended by properly adjusting the 
binary potential of the effective Hamiltonian, provided that the resonance 
is broad and entrance channel dominated \cite{KGG04}. We shall show
below that the 202 G resonance of $^{40}$K fulfils all 
requirements for an accurate treatment of the binary low energy bound and 
continuum spectrum in terms of just a single asymptotic scattering channel.

The single-channel approach of Refs.~\cite{KGB03,GKGTJ03,KGG04}  
effectively describes the resonance enhancement of the low energy 
scattering as a perturbation of the background scattering potential.
Following this approach, we replace the parameter $\xi_\mathrm{bg}$ in
Eq.~(\ref{separablepotentialgeneral}) by a magnetic field dependent amplitude
$\xi_\mathrm{eff-1ch}(B)$ and insert the associated effective single-channel 
potential 
\begin{align}
  V_\mathrm{eff-1ch}(B)=|\chi_\mathrm{bg}\rangle \xi_\mathrm{eff-1ch}(B)
  \langle\chi_\mathrm{bg}|
\end{align}   
into the general form of a single-channel Hamiltonian
\begin{align}
  H_\mathrm{2B-1ch}=
  -\frac{\hbar^2}{m}\nabla^2+V_\mathrm{eff-1ch}(B) \, .
  \label{H2B1channel}
\end{align}
In this treatment the range parameter $\sigma_\mathrm{bg}$ retains the 
same value as in the two-channel approach, while the amplitude 
$\xi_\mathrm{eff-1ch}(B)$ is adjusted in such a way that the effective 
potential recovers, at each value of the magnetic field, the scattering length 
$a(B)$ of Eq.~(\ref{aofB}). This adjustment can be performed in complete 
analogy to the determination of $\xi_\mathrm{bg}$ in the two-channel approach 
of Ref.~\cite{GKGTJ03}. The counterpart of Eq.~(\ref{condabg}) for the 
determination of $\xi_\mathrm{eff-1ch}(B)$ leads to the relationship:
\begin{equation}
  \xi_\mathrm{eff-1ch}(B)= \frac{4\pi \hbar^2
    a(B)/m}{1-a(B)/(\sqrt{\pi}\sigma_\mathrm{bg})} \, .
  \label{Veff1ch}
\end{equation} 
We note that despite the divergence of the scattering length $a(B)$ at
resonance the amplitude of the effective single-channel potential
$\xi_\mathrm{eff-1ch}(B)$ remains a smooth function of the magnetic field 
strength $B$. The single-channel approach not only recovers the exact 
magnetic field dependence of the scattering length, by identifying 
$a(B)$ in Eq.~(\ref{Veff1ch}) with Eq.~(\ref{aofB}), but it also leads to 
the exact asymptotic limit $E_{-1}$ of the binding energy $E_\mathrm{b}(B)$
away from the resonance (see Fig.~\ref{fig:E-1}), by using the same form 
factor $|\chi_\mathrm{bg}\rangle$ as the separable background scattering 
potential of Eq.~(\ref{separablepotentialgeneral}). 

%It can be shown \cite{GKGTJ03} that, within our
%2-channel formulation, very close to
%the zero-energy resonance the energy of the highest excited diatomic
%bound state satisfies the universal relation
%$E_\mathrm{b}=-\hbar^{2}/[m a^2]$. We note that the inclusion of
%the second parameter in the effective single channel description (the range 
%parameter $\sigma_\mathrm{bg}$) significantly extends the range of magnetic
%fields, where the proper dependence of $E_\mathrm{b}(B)$ is
%recovered. However, by its nature, even the range of validity of such a
%single channel approach is more limited than that of the two channel
%approach presented before.

\begin{figure}[!htbp]
  \begin{center}
    \includegraphics[width=\columnwidth,clip]{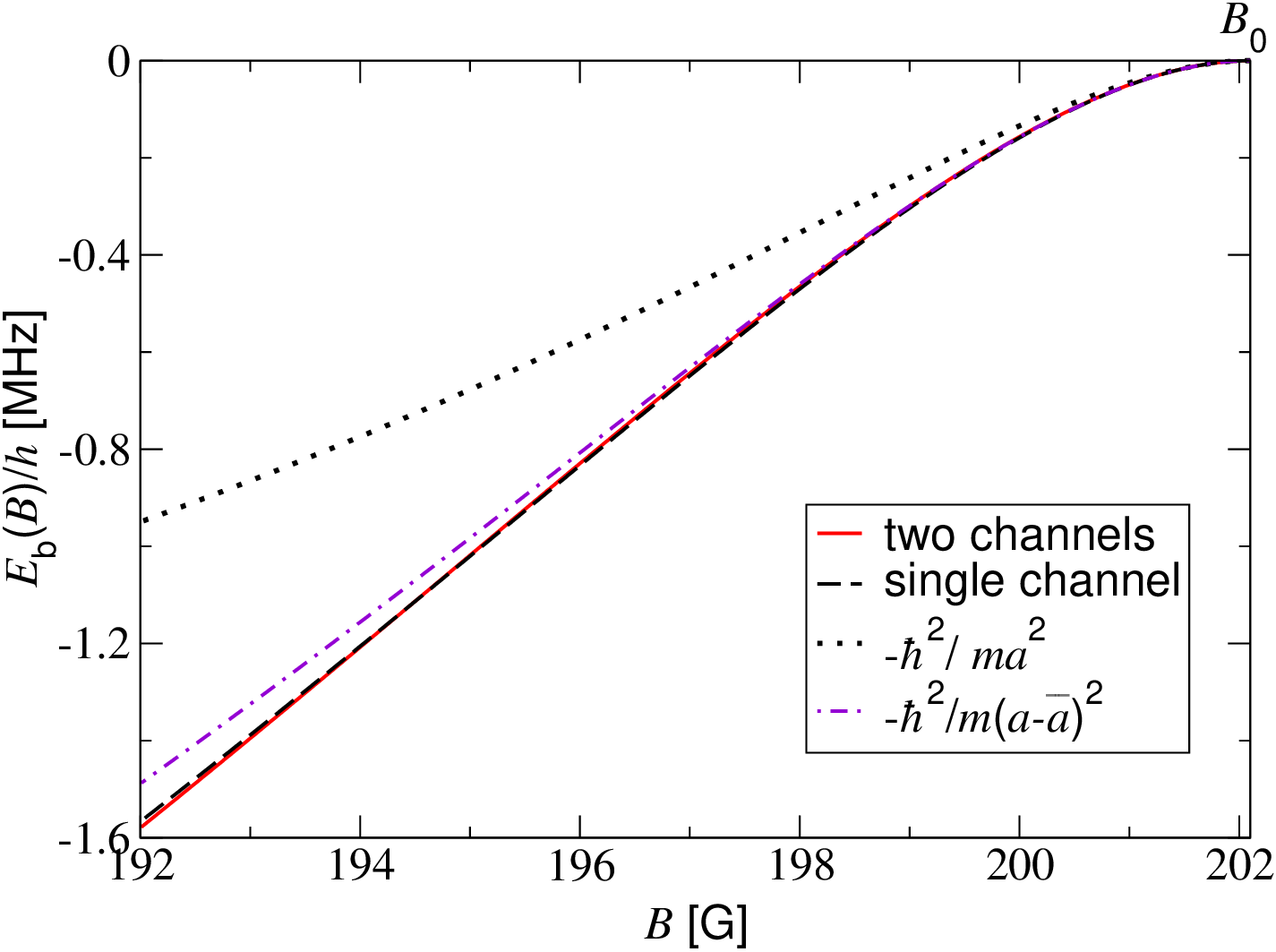}
    \caption{(Color online) Binding energy $E_\mathrm{b}(B)$ of the Fesh\-bach 
      molecular bound state $\phi_\mathrm{b}(B)$ associated with a pair of 
      unlike $^{40}$K atoms in the asymptotic $(f=9/2,m_f=-9/2)$ and 
      $(f=9/2,m_f=-7/2)$ Zeeman states versus the magnetic field strength
      $B$. The figure shows a comparison between the results of the
      two-channel approach (solid curve), the effective single-channel
      approach (dashed curve), and analytic estimates of the binding energy 
      (dotted and dot-dashed curves). While the dotted curve refers to the 
      universal estimate of Eq.~(\ref{Ebuniversal}), just depending on
      the scattering length $a(B)$, the dot-dashed curve also
      accounts for the non-retarded van der Waals interaction $-C_6/r^6$ of 
      the background scattering potential, at asymptotically large 
      inter-atomic distances $r$, in terms of the mean scattering
      length $\bar{a}$ \cite{Gribakin93,KGB03} via Eq.~(\ref{EbGF}).}
    \label{fig:40K}
  \end{center}
\end{figure}

Figure \ref{fig:40K} compares the binding energy $E_\mathrm{b}(B)$ of the 
Fesh\-bach molecule, as obtained from the two-channel approach of 
Ref.~\cite{GKGTJ03}, with the predictions of the effective single-channel 
approach, as well as analytic estimates. These estimates are based on 
Eq.~(\ref{Ebuniversal}) and on the formula 
\begin{align}
  E_\mathrm{b}\approx -\frac{\hbar^2}{m(a-\bar{a})^2},
  \label{EbGF}
\end{align}
which follows from the treatment of Ref.~\cite{Gribakin93} and accounts for 
the non-retarded van der Waals tail 
$V_\mathrm{bg}(r)\underset{r\to\infty}{\sim}-C_6/r^6$
of the background scattering potential, in terms of the mean scattering length 
$\bar{a}$ of Eq.~(\ref{abar}), in addition to the magnetic field dependence of 
the scattering length $a(B)$. As the magnetic field strength $B$ approaches 
the position $B_0$ of the zero-energy resonance from below, all predictions 
coincide. In this universal regime of magnetic field strengths the scattering 
length very much exceeds all the other length scales set by the binary 
interactions and completely determines the magnitude of the binding energy. 
Further away from the resonance the quality of the universal estimate of 
Eq.~(\ref{Ebuniversal}) deteriorates, while Eq.~(\ref{EbGF}) provides a good 
approximation of $E_\mathrm{b}(B)$ over the entire range of magnetic field 
strengths shown in Fig.~\ref{fig:40K}. The full predictions of the two- and 
single-channel approaches are virtually indistinguishable.
%and also in excellent agreement with multi-channel calculations
%\cite{Julienne}.
This agreement suggests that the admixture of the resonance level 
$\phi_\mathrm{res}$ to the Fesh\-bach molecule remains small also outside the 
universal regime of magnetic field strengths, i.e.~the 202 G resonance
of $^{40}$K is entrance channel dominated. In fact, we shall show in Section 
\ref{sec:Results} that the closed-channel admixture to the bound state wave 
function $\phi_\mathrm{b}(B)$ never exceeds about 8\% and reaches its maximum 
at about 7.2 G below the position $B_0=202.1$\,G of the zero-energy 
resonance.

%We conclude that, with regard to the energies of the weakly bound
%state  $\phi_\mathrm{b}(B)$ relevant to the experiments of
%Ref.~\cite{Regal04-2}, the effective single channel description should
%be sufficient, provided the parameter $\sigma_\mathrm{bg}$,
%accounting for the van der Waals part of the interaction potential,
%is properly taken into account.

\section{Many-body Hamiltonians}
\label{sec:model}
\subsection{The microscopic Hamiltonian}
Having established the microscopic parameters of the inter-atomic interactions,
which properly describe the experimentally relevant part of the two-body 
energy spectrum, we shall now discuss the many-body 
Hamiltonians that we use in our studies of the BCS-BEC cross-over in a 
cold dilute gas of $^{40}$K. The most general many-body Hamiltonian, including 
all atomic internal degrees of freedom, has the following form:
\begin{align}
  \nonumber
  \hat{H}_{\mathrm{MB}}=\sum_{\nu_\mathrm{1},\mathbf{k}} 
  \epsilon_{\mathbf{k},\nu_\mathrm{1}} 
  \hat{a}^{\dagger}_{\mathbf{k},\nu_\mathrm{1}}
  \hat{a}_{\mathbf{k},\nu_\mathrm{1}}+
  &\frac{1}{2} 
  \sum_{\nu_\mathrm{1}, \nu_\mathrm{2}, \nu_\mathrm{3}, \nu_\mathrm{4}} 
  \sum_{\mathbf{k},\mathbf{k}',\mathbf{q}}
  V_\mathrm{\nu_\mathrm{1}, \nu_\mathrm{2}, \nu_\mathrm{3}, \nu_\mathrm{4}}
  ({\mathbf{k}},{\mathbf{k}}')\\
  &\times
  \hat{a}^{\dagger}_{\frac{\mathbf{q}}{2}+{\mathbf{k}}, \nu_\mathrm{1}} 
  \hat{a}^{\dagger}_{\frac{\mathbf{q}}{2}-{\mathbf{k}},\nu_\mathrm{2}}
  \hat{a}_{\frac{\mathbf{q}}{2}-{\mathbf{k}}', \nu_\mathrm{3}} 
  \hat{a}_{\frac{\mathbf{q}}{2}+{\mathbf{k}}', \nu_\mathrm{4}} \, .
  \label{HMB}
\end{align}
Here the indices 
$\nu_\mathrm{1}$, $\nu_\mathrm{2}$, $\nu_\mathrm{3}$, $\nu_\mathrm{4}$ 
label the atomic Zeeman levels. The wave vector $\mathbf{k}$ in the kinetic 
energy contribution to Eq.~(\ref{HMB}) is associated with a single-particle 
momentum, while the pairs of wave vectors $\mathbf{q}$ and $\mathbf{k}$ in the 
interaction term correspond to two-body centre-of-mass and relative momenta. 
All these indices refer to our choice of plane wave basis states 
$\exp(i\mathbf{k}\cdot\mathbf{r})/\sqrt{\cal{V}}$ in a box of volume $\cal{V}$ 
with periodic boundary conditions. The associated annihilation and creation 
operators obey the usual fermionic anti-commutation relations:
\begin{align}
  \hat{a}_{\mathbf{k},\nu_\mathrm{1}}
  \hat{a}^\dagger_{\mathbf{k}',\nu_\mathrm{2}}
  +\hat{a}^\dagger_{\mathbf{k}',\nu_\mathrm{2}}
  \hat{a}_{\mathbf{k},\nu_\mathrm{1}}
  =&\delta_{\nu_\mathrm{1},\nu_\mathrm{2}}\delta_{\mathbf{k},\mathbf{k}'}\, ,\\
  \hat{a}_{\mathbf{k},\nu_\mathrm{1}}\hat{a}_{\mathbf{k}',\nu_\mathrm{2}}
  +\hat{a}_{\mathbf{k}',\nu_\mathrm{2}}
  \hat{a}_{\mathbf{k},\nu_\mathrm{1}}=&0\, .
\end{align}
We note that the microscopic potentials 
$V_\mathrm{\nu_\mathrm{1},\nu_\mathrm{2},\nu_\mathrm{3},\nu_\mathrm{4}}$
in Eq.~(\ref{HMB}) all decay to zero in the limit of large inter-atomic
separation. The atomic Zeeman energies, which determine the dissociation 
thresholds of the different two-body asymptotic scattering channels are 
included in the single-particle energies 
$\epsilon_{\mathbf{k},\nu_\mathrm{1}}$.

\subsection{Ambiguities between different many-body approaches} 
While a mean-field treatment of the BCS-BEC cross-over using 
realistic microscopic interactions for all binary scattering channels seems 
impractical, there are several ways to formulate approximate theories, 
derived from the Hamiltonian of Eq.~(\ref{HMB}) or the effective 
boson-fermion model \cite{Holland,Timm}, which are all compatible with 
the two-channel approach to the binary collision physics of 
Section \ref{sec:two-body}. One possible condition for such a compatibility 
may be, for instance, that the binding energy of the Fesh\-bach molecule, 
predicted by the two-channel approach, exactly coincides with twice the 
chemical potential obtained from the mean-field theory in the asymptotic 
limit of zero density. 

During the course of our studies we have derived such a mean-field approach 
\cite{Szymanska04}, based on the microscopic Hamiltonian of Eq.~(\ref{HMB}) 
and the general kinetic theory of Refs.~\cite{Fricke96,KB02}, by applying
the two-channel approach of Section \ref{sec:two-body}, including the 
single-resonance approximation, to the two-body propagators. This treatment 
is most sensible as it relies upon controlled approximations rather than 
uncontrollable model assumptions about the form of an effective Hamiltonian. 
The dynamic equations for the one-body density matrix, however, explicitly 
depend on the single-atom Zeeman states that constitute the closed channel 
associated with the Fesh\-bach resonance level. The atomic energy levels 
need to be explicitly specified in such a description once one and the same 
single-particle Zeeman state is shared between the relevant two-body 
entrance and closed channels. This phenomenon has been discussed 
previously in Ref.~\cite{meera} for the case of $^{40}$K and reflects physical 
intuition, since measurements of single-particle quantities, like the position 
of an atom or its internal state, do not necessarily provide information about 
whether this atom was correlated with another partner in the spatial 
configuration of the Fesh\-bach resonance level.

The assumption of such a generally unphysical possibility of distinguishing 
spatially separated from correlated atom pairs in a gas on the basis of 
one-body observables underlies the boson-fermion model 
\cite{Holland,Timm}. This model separates out the Fesh\-bach resonance state 
on the level of the Hamiltonian. We shall show, however, that the standard
thermodynamic physical quantities we discuss in Section \ref{sec:Results} 
are recovered, at least on the level of the mean-field approximation, by 
many-body approaches based on both the boson-fermion and the standard 
fermionic Hamiltonians, provided that they are suitably adjusted to be 
compatible with the precise two-channel approach to the two-body physics 
of Section \ref{sec:two-body}. This observation suggests that the thermal 
equilibrium properties of Section \ref{sec:Results} are insensitive to the 
underlying model Hamiltonian, regardless of the level of accuracy to which 
they include the separation into different asymptotic binary scattering 
channels.

\subsection{Adjustment of the boson-fermion model} 
The boson-fermion model leads to an effective two-channel 
description of the two-body physics and includes only the interactions 
between unlike fermions in the different Zeeman states associated with 
the entrance channel of Section \ref{sec:two-body}. The boson-fermion
Hamiltonian has the following form:
\begin{align}
  \nonumber
  \hat{H}_{\mathrm{MB-BF}}&=\sum_{s,\mathbf{k}} \epsilon_{\mathbf{k}}
  \hat{a}^{\dagger}_{\mathbf{k},s}\hat{a}_{\mathbf{k},s}+ \sum_{\mathbf{q}} 
  \left[E_{\mathbf{q}}+
  E_\mathrm{res}(B)\right]\hat{b}^{\dagger}_{\mathbf{q}}\hat{b}_{\mathbf{q}} \\
  \nonumber
  &+\sum_{\mathbf{k},\mathbf{k'},\mathbf{q}} 
  V_\mathrm{bg}({\mathbf{k}},{\mathbf{k}}') 
  \hat{a}^{\dagger}_{\frac{\mathbf{q}}{2}+{\mathbf{k}}, \uparrow}
  \hat{a}^{\dagger}_{\frac{\mathbf{q}}{2}-\mathbf{k},\downarrow} 
  \hat{a}_{\frac{\mathbf{q}}{2}-{\mathbf{k}}',\downarrow} 
  \hat{a}_{\frac{\mathbf{q}}{2}+{\mathbf{k}}',\uparrow}
  \\ 
  &+\sum_{\mathbf{k},\mathbf{q}} \
  g({\mathbf{k}})\left(\hat{b}^{\dagger}_{\mathbf{q}} 
  \hat{a}_{\frac{\mathbf{q}}{2}-\mathbf{k},\downarrow}
  \hat{a}_{\frac{\mathbf{q}}{2}+{\mathbf{k}},\uparrow} + h.c.\right) \, .
  \label{HMB2ch}
\end{align}
Here $\hat{b}_{\mathbf{q}}$ and $\hat{b}^{\dagger}_{\mathbf{q}}$ annihilate or 
create pairs of atoms with the centre-of-mass momentum $\mathbf{q}$ in the 
Fesh\-bach resonance level, respectively, and fulfil the bosonic commutation 
relations:
\begin{align}
  \hat{b}_{\mathbf{q}}\hat{b}^{\dagger}_{\mathbf{q}'}-
  \hat{b}^{\dagger}_{\mathbf{q}'}\hat{b}_{\mathbf{q}}
  &=\delta_{\mathbf{q},\mathbf{q}'}\, ,\\
  \hat{b}_{\mathbf{q}}\hat{b}_{\mathbf{q}'}-
  \hat{b}_{\mathbf{q}'}\hat{b}_{\mathbf{q}}
  &=0\, .
\end{align}
The kinetic energy associated with the centre of mass of an atom pair in
the meta-stable Fesh\-bach resonance level is denoted by $E_{\mathbf{q}}$.
In analogy to related applications in the theory of superconductivity, we 
choose, in the following, the notation $s=\uparrow,\downarrow$ to label 
the single-particle Zeeman levels that constitute the $s$-wave entrance 
channel. We assume all fermionic operators $\hat{a}_{\mathbf{k},s}$ and 
$\hat{a}^{\dagger}_{\mathbf{k},s}$ to commute with all bosonic operators 
$\hat{b}_{\mathbf{q}}$ and $\hat{b}^{\dagger}_{\mathbf{q}}$. In analogy to our 
considerations of the two-body problem in Section \ref{sec:two-body} we have 
included the magnetic field dependent single-particle Zeeman energies in the 
binary interactions. The single-particle energies $\epsilon_{\mathbf{k}}$
are therefore independent of the index $s$ associated with atomic Zeeman 
levels.

In order to determine the parameters of the boson-fermion Hamiltonian,
we compare the Schr\"odinger equations obtained from Eq.~(\ref{HMB2ch}) and 
the binary two-channel Hamiltonian of Eq.~(\ref{H2B2channel}) when applied
to a general two-body state with components in both the entrance channel and 
the Fesh\-bach resonance level. As both approaches are required to describe
the same physics, this comparison shows that $E_\mathrm{res}(B)$ is the
resonance energy given by Eq.~(\ref{Eres}), while $V_\mathrm{bg}$ is the 
background scattering potential of Eq.~(\ref{separablepotentialgeneral}). 
Its matrix element in terms of the energy states of the periodic box is thus 
given by
\begin{align}
  V_\mathrm{bg}({\mathbf{k}},{\mathbf{k}}')=\frac{1}{\mathcal{V}}
  \int_{\mathcal{V}} d\mathbf{r}\, \int_{\mathcal{V}} d\mathbf{r'}\, 
  e^{-i\mathbf{k}\cdot\mathbf{r}}
  V_\mathrm{bg}({\mathbf{r}},{\mathbf{r}}')
  e^{i\mathbf{k}'\cdot\mathbf{r}'}\,,
\end{align}
where the integration extends over the volume $\mathcal{V}$ of the box.
The parameter 
\begin{align}
  g({\mathbf{k}})=\frac{1}{\sqrt{\mathcal{V}}}
  \int_{\mathcal{V}} d\mathbf{r}\, e^{-i\mathbf{k}\cdot\mathbf{r}}
  W(r)\phi_\mathrm{res}(r)\, , 
\end{align}
which characterises the inter-channel coupling, can be obtained from 
Eq.~(\ref{coupling}). We note that with these adjustments of potentials and
parameters the boson-fermion Hamiltonian of Eq.~(\ref{HMB2ch}) and the 
two-channel approach in the single resonance approximation of Section 
\ref{sec:two-body} provide exactly identical descriptions of the two-body 
physics.

\subsection{Adjustment of the standard fermionic Hamiltonian}
A many-body approach compatible with the single-channel description of 
resonance enhancement in the binary low energy collision physics of Section 
\ref{sec:two-body} can be obtained from a many-body Hamiltonian usually 
applied in the theory of superconductivity 
\cite{Leggett,Noz-SchR,Schrieffer}. This standard fermionic Hamiltonian is 
given by the following expression: 
\begin{align}
  \nonumber
  \hat{H}_{\mathrm{MB-F}}=\sum_{s,\mathbf{k}} \epsilon_{\mathbf{k}}
  \hat{a}^{\dagger}_{\mathbf{k},s} \hat{a}_{\mathbf{k},s}&+
  \sum_{\mathbf{k},\mathbf{k}',\mathbf{q}} 
  V_\mathrm{eff-1ch}({\mathbf{k}},{\mathbf{k}}',B)\\
  &\times
  \hat{a}^{\dagger}_{\frac{\mathbf{q}}{2}+{\mathbf{k}},\uparrow} 
  \hat{a}^{\dagger}_{\frac{\mathbf{q}}{2}-{\mathbf{k}},\downarrow} 
  \hat{a}_{\frac{\mathbf{q}}{2}-{\mathbf{k}}', \downarrow} 
  \hat{a}_{\frac{\mathbf{q}}{2}+{\mathbf{k}}',\uparrow} \, .
  \label{HMB1ch}
\end{align}
Here $V_\mathrm{eff-1ch}(B)$ is the interaction potential for a pair of atoms 
in the Zeeman states associated with the entrance channel, which we have 
labelled by the indices $s=\uparrow,\downarrow$. We choose this potential to 
be identical to the binary interaction in the single-channel Hamiltonian of 
Eq.~(\ref{H2B1channel}). This choice assures that the two-body physics 
described by the standard fermionic Hamiltonian exactly agrees with the 
single-channel approach of Section \ref{sec:two-body}.

As both the two- and single-channel approaches of Section \ref{sec:two-body}
yield virtually the same low energy two-body physics, we may also expect a 
similar agreement between the predictions obtained from the Hamiltonians of 
Eqs.~(\ref{HMB2ch}) and (\ref{HMB1ch}) with regard to the thermal equilibrium 
physical quantities of Section \ref{sec:Results}. The formal establishment of 
this equivalence of these many-body Hamiltonians in applications to 
standard mean-field theory will be the subject of the following sections. 
Since we have included smooth binary potentials with a proper spatial extent 
in the Hamiltonians, our mean-field calculations do not involve ultra-violet 
divergences and, therefore, do not require any renormalisation procedures.

\section{Mean-field approach to the thermal equilibrium} 
\label{sec:therm}

At temperatures close to absolute zero or well below the critical 
temperature $T_\mathrm{c}$, the BCS-BEC cross-over has been successfully 
analysed by mean-field theory in terms of the BCS wave function, which 
smoothly interpolates between the BCS and BEC regimes 
\cite{Leggett,Keldysh,Rand}. The derivation of the thermal equilibrium 
properties \cite{Ohashi02,Milstein02,Perali03,Combescot03,Viverit04-1,Stoof,Mackie05,Bruun04-1,Carr04,Karpiuk04,Perali04-1,Stajic04,DePalo04,Heiselberg04,Buchler04,Perali04-2,Kinnunen04,Williams04,Drummond04,Bulgac04,Diener04-1,Avdeenkov05,Parish05,Diener04-2,Jensen04} follows standard techniques (see, e.g., Ref.~\cite{Schrieffer}). 
We thus include just a brief derivation of their extension to the 
boson-fermion model, applying path integral methods \cite{Negele98}. 
The associated mean-field approach for the standard fermionic Hamiltonian 
model can be obtained from our analysis simply by replacing 
$g(\mathbf{k})\to 0$ and $V_{\mathrm{bg}}\to V_{\mathrm{eff-1ch}}$ in all 
formulae.

\subsection{Path integral approach}
All thermodynamic equilibrium physical quantities of a gas can be derived 
from the quantum partition function, which is given by the general formula:
\begin{align}
  \mathcal{Z}=\mathrm{tr}
  \left[ e^{-\beta \left(\hat{H}_\mathrm{MB-BF}-\mu \hat{N}\right)}\right]. 
\end{align}
Here the factor $\beta=1/(k_\mathrm{B}T)$ accounts for the temperature
$T$ scaled to energy units by the Boltzmann constant $k_\mathrm{B}$,
$\hat{N}$ is the number operator and $\mu$ the chemical potential,
while ``$\mathrm{tr}$'' indicates the trace of an operator. The quantum
partition function $\mathcal{Z}$ can be conveniently represented in
terms of a coherent state path integral. To this end, we introduce the
action
\begin{equation}
  S=\int_0^\beta d\tau 
  \left(
  H_{\mathrm{MB-BF}} - \sum_{\mathbf{q}} \ 
  \bar{b}_{\mathbf{q}}\partial_\tau b_{\mathbf{q}} - 
  \sum_{\mathbf{k}} \bar{\psi}_\mathbf{k}
  \partial_\tau \psi_\mathbf{k}- \mu N
  \right) \, ,
  \label{action}
\end{equation}
where $N$ is the total number of atoms and $H_{\mathrm{MB-BF}}$ is the
many-body boson-fermion Hamiltonian 
of Eq.~(\ref{HMB2ch}). In this Hamiltonian $\hat{a}_{\mathbf{k},s}$ 
and $\hat{b}_\mathbf{q}$ are replaced by $\tau$-dependent Grassmann numbers 
$a_{\mathbf{k},s}$ and c-numbers $b_\mathbf{q}$, respectively. We
have, furthermore, introduced the quantities 
$\bar{a}_{\mathbf{k},s}$, $\bar{b}_{\mathbf{q}}$ as well as the Nambu spinors
\begin{align}
  \psi_{{\mathbf{k}}}=\left(
  \begin{array}{c}
    a_{\mathbf{k},\uparrow}\\
    \bar{a}_{-\mathbf{k},\downarrow}
  \end{array}
  \right)\, ,\,
  \bar{\psi}_{{\mathbf{k}}}= (\bar{a}_{\mathbf{k},\uparrow}, 
      {a}_{-\mathbf{k},\downarrow}) 
\end{align}
in Eq.~(\ref{action}), where the bar indicates 
that the quantities $\bar{a}_{\mathbf{k},s}$ and $a_{\mathbf{k},s}$, etc., are 
independently varied. In our considerations $\bar{a}_{\mathbf{k},s}$ and 
$a_{\mathbf{k},s}$ as well as $b_{\mathbf{q}}$ and $\bar{b}_{\mathbf{q}}$ are 
all related by complex conjugation. The coherent state path integral then 
determines quantum partition function $\mathcal{Z}$ in terms of the action to 
be: 
\begin{equation}
  \mathcal{Z}=\int D(\bar{\psi},\psi,\bar{b},b)e^{-S} \, .
  \label{pathintegral}
\end{equation}    

\subsection{Relationship between the path integral approach and the
  zero-temperature mean-field approximation}

In the following, we shall briefly summarise the assumptions underlying our 
evaluation of Eq.~(\ref{pathintegral}) in the mean-field approximation as well 
as their relationship to derivations based on the condensate wave function of 
Refs.~\cite{Leggett,Keldysh,Rand}. In order to treat the interaction we 
isolate the components associated with the pairing. Such a procedure is 
equivalent to considering only those contributions of the total Hamiltonian 
that yield non-zero matrix elements in the condensate state, i.e., 
$\langle \Phi |\hat{H}_{\mathrm{MB-BF}}|\Phi\rangle\ne0$.  
Following the ideas of standard BCS theory \cite{Schrieffer}, 
the associated zero-temperature ground state condensate wave function is 
given by the formula: 
\begin{align}
  \nonumber
  |\Phi \rangle &=
  \frac{
    \exp
    \left(-|\lambda_\mathrm{b}|^2/2
    +{\lambda_\mathrm{b}}\hat{b}^\dag_{0}
    +\sum_{{\mathbf{k}}} 
    w_{{\mathbf{k}}}
    \hat{a}^\dag_{{{\mathbf{k}}},\uparrow}
    \hat{a}^\dag_{{-\mathbf{k}},\downarrow} 
    \right)}
       {\prod_{\mathbf{k}'}
	 \left(1+|w_{\mathbf{k}'}|^2\right)^{1/2}}
       |\mathrm{vac}\rangle\\
       &=
       e^{-|\lambda_\mathrm{b}|^2/2+{\lambda_\mathrm{b}}\hat{b}^\dag_{0}} 
       \prod_{{\mathbf{k}}} \left(u_{{\mathbf{k}}}
       + v_{{\mathbf{k}}} \hat{a}^\dag_{{{\mathbf{k}}},\uparrow}
       \hat{a}^\dag_{{-\mathbf{k}},\downarrow}\right) |\mathrm{vac}\rangle\,.
       \label{condensatewavefunction}
\end{align}
Here $|\mathrm{vac}\rangle$ is the vacuum state. We have, furthermore, 
introduced the quantity $w_{\mathbf{k}}=v_{\mathbf{k}}/u_{\mathbf{k}}$
in Eq.~(\ref{condensatewavefunction}) which depends on the quasi-particle 
amplitudes $u_{\mathbf{k}}$ and $v_{\mathbf{k}}$. The quasi-particle 
amplitudes fulfil the requirement
$|u_{\mathbf{k}}|^2+|v_{\mathbf{k}}|^2=1$, where
\begin{equation}
  2\left|
  v_{\mathbf{k}}
  \right|^2=
  \langle\Phi|
  \left(
  \hat{a}^\dag_{{{\mathbf{k}}},\uparrow}
  \hat{a}_{{\mathbf{k}},\uparrow}
  +\hat{a}^\dag_{{{\mathbf{k}}},\downarrow}
  \hat{a}_{{\mathbf{k}},\downarrow}
  \right)
  |\Phi\rangle
\end{equation}
determines the average occupation number of the momentum mode associated with 
the wave vector $\mathbf{k}$. The product 
$\kappa_\mathbf{k}=u_{\mathbf{k}}v_{\mathbf{k}}$ can be interpreted as a pair 
wave function, whose modulus squared, summed over all wave vectors 
$\mathbf{k}$ (i.e.~$\sum_\mathbf{k}|u_{\mathbf{k}}v_{\mathbf{k}}|^2$),
provides a measure for the number of condensed fermionic pairs. Within the 
framework of the boson-fermion model, the coefficient $\lambda_\mathrm{b}$ in 
Eq.~(\ref{condensatewavefunction}) is related to the average number of pairs 
of atoms in the resonance state configuration through the formula:
\begin{equation}
  \left|\lambda_\mathrm{b}\right|^2=
  \langle\Phi|\hat{b}_0^\dag\hat{b}_0|\Phi\rangle\,.
\end{equation} 

We shall omit, in our evaluation of Eq.~(\ref{pathintegral}), the decoupling 
in the diagonal, Hartree-Fock, channel because, firstly, the Fock pairing, 
which results from exchange, requires indistinguishable atoms in the same 
Zeeman states. The associated binary interactions, however, which all involve 
non-isotropic partial waves, are omitted from the outset in the $s$-wave 
boson-fermion Hamiltonian. Secondly, the inclusion of the Hartree term
results in only a negligible energy shift. 
%In the excitonic BCS-BEC cross-over problem or in applications to 
%traditional superconductors this energy shift is counterbalanced by the 
%positive background.

\subsection{Hubbard-Stratonovich transformation}
To further evaluate the finite temperature partition function of 
Eq.~(\ref{pathintegral}), we proceed by applying the Hubbard-Stratonovich 
decoupling to the interaction in the off-diagonal channel, which yields: 
\begin{widetext}
\begin{align}  
  \mathcal{Z}=\int D(\bar{\psi},\psi,\bar{b}_0,b_0) \int
  D(\Delta^*,\Delta)
  \exp\Bigg[
    \int_0^\beta d\tau 
    \sum_{\mathbf{k}} \ \bar{\psi}_{{{\mathbf{k}}}} 
    \begin{pmatrix} 
      \partial_\tau -\epsilon_{{\mathbf{k}}}+\mu  &
      -g({\mathbf{k}})b_0 -\Delta_{{\mathbf{k}}} \\
      -g({\mathbf{k}})\bar{b}_0-\Delta^*_{{\mathbf{k}}} &
      \partial_\tau +\epsilon_{{\mathbf{k}}}-\mu 
    \end{pmatrix} \psi_{{{\mathbf{k}}}}  -S_0 
    \Bigg]\, .
  \label{ZHubbardStratonovich}
\end{align} 
\end{widetext}
The transformed partition function depends on the order parameter
$\Delta_\mathbf{k}$, sometimes referred to as the Hubbard-Stratonovich
field, whose explicit expression within the mean-field approximation
in the limit of zero temperature reads
$\Delta_\mathbf{k}=-\sum_{\mathbf{k}'}
V_\mathrm{bg}({\mathbf{k}},{\mathbf{k}'}) \langle\Phi|
\hat{a}_{-\mathbf{k}',\downarrow}
\hat{a}_{\mathbf{k}',\uparrow}|\Phi\rangle$.  In our more general
finite temperature treatment the order parameter is a $\tau$-dependent
function which we shall determine using a saddle-point analysis. We
have, furthermore, introduced in Eq.~(\ref{ZHubbardStratonovich}) the
quantity
\begin{align}
  \nonumber
  S_0&=\int_0^\beta d\tau\, \sum_{\mathbf{k},\mathbf{k}'}
  \Delta^*_{{\mathbf{k}}} V^{-1}_\mathrm{bg}({\mathbf{k}},{\mathbf{k}'})
  \Delta_{{\mathbf{k}'}}\\ 
  &-\int_0^\beta d\tau\,
  \bar{b}_0\left[\partial_\tau-E_{\mathrm{res}}(B)+2\mu\right]b_0\, ,
\label{S0}
\end{align}
where $V^{-1}_\mathrm{bg}$ indicates the inverse potential energy operator
associated with the background scattering in the relative motion of a 
pair of unlike fermions. We note that, in the mean-field approximation, the 
partition function of Eq.~(\ref{ZHubbardStratonovich}) depends only 
on the c-numbers $b_{\mathbf{q}=0}$ and $\bar{b}_{\mathbf{q}=0}$
associated with the zero-momentum mode of the centre-of-mass motion of a
pair of fermions in the resonance level.
 
Since the exponent in Eq.~(\ref{ZHubbardStratonovich}) is a quadratic form 
in the fermionic fields, the associated Gaussian integral can be performed
straightforwardly. This integration leads to the following formula:
\begin{equation}
  \mathcal{Z}= \int D(\Delta^*,\Delta,\bar{b}_0,b_0) e^{-S_0} \exp\left(
    \mathrm{tr}\left[\ln G^{-1}\right]\right)\, .
  \label{part}
\end{equation}
This formula depends on the thermal Green's function $G$, whose inverse 
operator is given by: 
\begin{align}
  \nonumber
  G^{-1}=&\partial_\tau-\left(\epsilon_{{\mathbf{k}}}-\mu\right)
  \sigma_3-\left[g({\mathbf{k}})b_0+\Delta_{{\mathbf{k}}}\right]\sigma_+
  \\
  &
  -\left[g({\mathbf{k}})\bar{b}_0+\Delta^*_{{\mathbf{k}}}\right]\sigma_-\,.
\label{G-1}
\end{align}
Here $\sigma_1$, $\sigma_2$ and $\sigma_3$ are the usual two-dimensional 
Pauli spin matrices, i.e.~their commutation relations are given by 
$\sigma_1\sigma_2-\sigma_2\sigma_1=2i\sigma_3$ and its cyclic permutations, 
while $\sigma_+=(\sigma_1+i\sigma_2)/2$ and $\sigma_-=(\sigma_1-i\sigma_2)/2$ 
are the associated raising and lowering matrices. 

\subsection{Saddle-point analysis}
%We shall assume in the following that the Fourier components of the
%order parameters $\Delta_\mathbf{k}(\tau)$, with respect to the
%variable $\tau$, are sharply peaked only at twice the chemical
%potential. 

We shall apply, in the following, a saddle-point analysis in which
the order parameters $\Delta_\mathbf{k}(\tau)$ and $b_0(\tau)$ are
assumed to be independent of $\tau$. This approach minimises the free energy 
as the derivatives of $\Delta_\mathbf{k}$ and $b_0$, with respect to $\tau$, 
do not contribute. The Green's function $G$ is then diagonal in the Matsubara 
frequencies and in the wave vectors ${\mathbf{k}}$. We shall, therefore, 
employ a discrete Fourier analysis to change the representation from the
variable $\tau$ to the Matsubara frequencies $\omega_j=(2j+1)\pi/\beta$,
where $j$ is an integer.  Performing the Fourier expansion of
$G^{-1}$, after inversion, the mean-field thermal Green's function is
given by the formula:
\begin{align}
  \nonumber
  G_{{{\mathbf{k}}}\omega_j}=\frac{1}{\omega_j^2+E_{{\mathbf{k}}}^2}
  \Big\{
    &
    -i\omega_j-(\epsilon_{{\mathbf{k}}}-\mu)\sigma_3 -
    \left[g({\mathbf{k}})b_0+\Delta_{{\mathbf{k}}}\right]\sigma_+ \\
    &
    -\left[g({\mathbf{k}})\bar{b}_0+\Delta^*_{{\mathbf{k}}}\right]\sigma_- 
    \Big\}\, .
\end{align}
Here $E_{{\mathbf{k}}} = \sqrt{(\epsilon_{{\mathbf{k}}}-\mu)^2+
|g({\mathbf{k}})b_0+\Delta_{{\mathbf{k}}}|^2}$ are the single-particle 
excitation energies. Varying the action with respect to 
$\Delta_{{\mathbf{k}}}^*$ and $\bar{b}_0$ leads to the saddle-point 
(mean-field) equations. 
%
%\begin{eqnarray*}
%\sum_{\mathbf{k}'} V^{-1}_{bg}({\mathbf{k}},{\mathbf{k}'})
%\Delta_{\mathbf{k}'}&=&
%\sum_{\omega_n}
%\frac{g({\mathbf{k}})b_0+
%\Delta_{{\mathbf{k}}}}{\omega_n^2+E_{{\mathbf{k}}}^2}\\
%(-E_{\mathrm{res}}+2\mu)b_0 &=& \sum_{{\mathbf{k}}}g({\mathbf{k}})
%\sum_{\omega_n}
%\frac{g({\mathbf{k}})b_0+
%\Delta_{{\mathbf{k}}}}{\omega_n^2+E_{{\mathbf{k}}}^2}.
%\end{eqnarray*}
%
We then perform the standard Matsubara frequencies summation and
consider the uniform gas limit, obtained by the following substitution
$\sum_{\mathbf{k}} \to \int
{\mathcal{V}}\frac{d\mathbf{k}}{(2\pi)^3}=\int {\cal{V}} 
\frac{d\mathbf{p}}{(2\pi\hbar)^3}$.
%, where $p$ is a wave-vector normalised to incorporate the
%$(2\pi\hbar)^3$ factor as in (\ref{separablepotentialGaussian}).
In the following, we shall use the explicit form of the potentials of Section
\ref{sec:two-body}, i.e.~the background scattering potential 
$V_{\mathrm{bg}}(p,p')=\xi_{\mathrm{bg}}
\chi_{\mathrm{bg}}(p)\chi_{\mathrm{bg}}(p')$ and the inter-channel coupling
$g(p)=\zeta \chi(p)$, in addition to the abbreviation 
\begin{equation}
  \int d\mathbf{p}\, \chi_{\mathrm{bg}}(p)
  \Delta_{p} = \xi_{\mathrm{bg}}\Delta\,. 
\end{equation}
The saddle-point equations are then given by the following pair of formulae:
\begin{align}
  \label{sp1}
  &\Delta = \int d\mathbf{p} \frac{\chi_{\mathrm{bg}}(p)\Sigma(p)} 
	{{2\sqrt{(\epsilon_p-\mu)^2+|\Sigma(p)|^2}}} 
	\mathrm{tanh}\left(\frac{\beta E_p}{2}\right),\\
	&\left[E_{\mathrm{res}}(B)-2\mu\right]b_0 = \int d\mathbf{p}
	\frac{\zeta \chi(p)\Sigma(p)}{2\sqrt{(\epsilon_p-\mu)^2+
	    |\Sigma(p)|^2}} \mathrm{tanh}\left(\frac{\beta E_p}{2}\right).
	\label{sp2}
\end{align}
Here the quantity 
\begin{equation}
  \Sigma(p)=\zeta \chi(p) b_0 + \xi_{\mathrm{bg}} \chi_{\mathrm{bg}}(p)
  \Delta 
\end{equation}
may be interpreted as a self energy.
The chemical potential $\mu$ is determined by the density equation
\begin{equation}
  n=\frac{1}{(2\pi \hbar)^3}\left[2\bar{b}_0 b_0+ \int d\mathbf{p}\, \left(1-
    \frac{\epsilon_{p}-\mu}
	 {{{\sqrt{(\epsilon_{p}-\mu)^2+ |\Sigma(p)|^2}}}} \right) \right]\, ,
  \label{dens}
\end{equation}
where $n$ is the number of atoms per unit volume. The mean-field
partition function reads: 
\begin{equation}
  \mathcal{Z}_{\mathrm{MF}}=  e^{-S_0} \exp\left(
    \mathrm{tr}\left[\ln G^{-1}\right]\right)\, .
  \label{part-fin}
\end{equation}
Here $S_0$ and $G^{-1}$, given by Eqs.~(\ref{S0}) and (\ref{G-1}),
respectively, are now evaluated at the saddle point, i.e.~$b_0$ and
$\Delta$ are the solutions of Eqs.~(\ref{sp1}), (\ref{sp2}) and
(\ref{dens}).
\subsection{Solutions of the saddle-point equations}
In Section \ref{sec:Results} we provide the numerical solutions of the
three coupled equations (\ref{sp1}), (\ref{sp2}) and (\ref{dens}). To
obtain a qualitative overview of their behaviour, we note that in the
zero-density limit $n\to 0$ the chemical potential $\mu$ recovers half
the molecular binding energy and the equations (\ref{sp1}),
(\ref{sp2}) and (\ref{dens}) reduce to the two-body Schr\"odinger
equation associated with the two-channel Hamiltonian
(\ref{H2B2channel}).  There are, however, two different solutions to
this equation. One of them corresponds to the weakly bound Fesh\-bach
molecular state just below the entrance channel dissociation threshold
(solid curve in Figure \ref{fig:E-1}). This state can be efficiently
populated by an adiabatic sweep of the magnetic field strength. The
second solution corresponds to the next, more tightly bound, molecular
state of energy given by the dashed curve in Figure
\ref{fig:E-1}. Similarly, at the many-body level (i.e.~$n \neq 0$)
there are three different extrema of the effective action in
Eq.~(\ref{part}), which are all solutions of Eqs.~(\ref{sp1}),
(\ref{sp2}) and (\ref{dens}) associated with different chemical
potentials. The trivial uncondensed solution ($\Delta=0$, $b_0=0$) is
unstable below $T_\mathrm{c}$, but there are two stable solutions with
a finite condensate component. One of them corresponds to the
condensation of weakly bound Fesh\-bach molecules with an energy just
below the entrance channel dissociation threshold (solid curve in
Figure \ref{fig:E-1}).  The other solution corresponds to the
condensation of molecules in the more tightly bound state (dashed
curve in Figure \ref{fig:E-1}). The latter solution is given by a
lower minimum of the effective action in Eq.~(\ref{part}), but it is
not populated on the time scale of the experiments of
Ref.~\cite{Regal04-2}. Instead, the local minimum (quasi-equilibrium)
solution is the physical one, relevant to the experiments of
Ref.~\cite{Regal04-2}. We shall analyse this solution in Section
\ref{sec:Results}.

The saddle-point equations for the mean-field approach associated
with the standard fermionic Hamiltonian of Eq.~(\ref{HMB1ch}) can be easily 
obtained from Eqs.~(\ref{sp1}), (\ref{sp2}) and (\ref{dens}) in the limit 
$g(p)\to 0$, in addition to the replacement 
$V_{\mathrm{bg}}\to V_{\mathrm{eff-1ch}}$. The 
single-channel effective potential $V_{\mathrm{eff-1ch}}$ supports just a 
single bound state, the Fesh\-bach molecule, which is efficiently populated in 
the experiments of Ref.~\cite{Regal04-2}. The associated mean-field equations,
therefore, have only two solutions: the uncondensed one, unstable below 
$T_\mathrm{c}$, and the condensate of weakly bound molecules, which we shall
analyse in Section \ref{sec:Results}.

\section{BCS-BEC cross-over in a dilute $^{40}$K gas}
\label{sec:Results}
Using the microscopic parameters determined in Section
\ref{sec:two-body}, we have numerically solved the coupled saddle-point
equations (\ref{sp1}), (\ref{sp2}) and (\ref{dens}) for a wide range of 
magnetic fields and atomic densities. In particular, we have considered 
magnetic field strengths from 1 G above to 10 G below the position of the 
zero-energy resonance ($B_0=202.1$ G \cite{Regal04-2}) in a mixture of 
fermionic $^{40}$K atoms in $(f=9/2,m_f=-9/2)$ and $(f=9/2,m_f=-7/2)$ 
Zeeman states. This range covers all magnetic field strengths relevant 
to Ref.~\cite{Regal04-2} and includes both the universal regime
around the zero-energy resonance (extending to about 1 G below $B_0$) and 
the region in which the energy of the Fesh\-bach molecule is strongly 
influenced by the anti-crossing between the energies of the two highest 
excited diatomic multi-channel vibrational bound states (solid and dashed 
curves in Fig.~\ref{fig:E-1}). 
We have also considered a wide range of experimentally accessible 
densities from $10^{12}$ cm$^{-3}$ to $5\times10^{14}$ cm$^{-3}$. In order to 
establish a characteristic value of the density of atoms in a homogeneous 
two-component gas, we require it to reproduce the Fermi energy of the trapped 
gas for the conditions of Ref.~\cite{Regal04-2}. This characteristic density 
is equal to $1.5\times 10^{13}$ cm$^{-3}$ and is given by the two-component 
peak density of atoms in the trap.

\subsection{Standard fermionic versus boson-fermion approach}
We have performed the numerical analysis for both approaches, using 
the standard fermionic Hamiltonian and the boson-fermion Hamiltonian, and 
found that in the entire range of experimentally relevant magnetic field 
strengths and densities there is hardly any difference visible between them  
for the standard many-body observables considered in this paper. 
Consequently, in the following discussion, we just give the results obtained 
from the standard fermionic Hamiltonian and explicitly point out small 
differences with respect to the boson-fermion approach when they arise.

\begin{figure}[htbp]
  \centering
  \includegraphics[width=\columnwidth,clip]{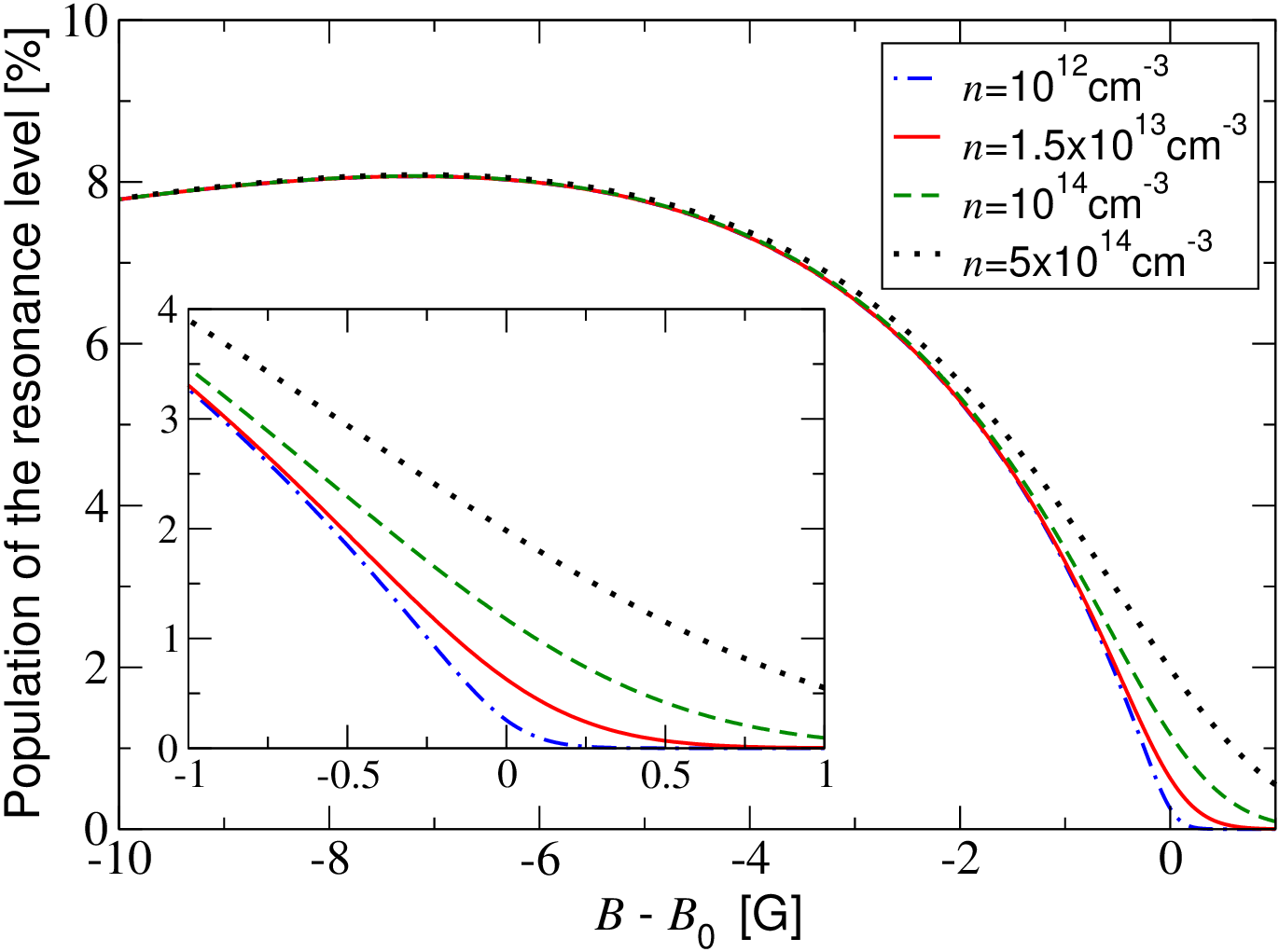}
  \caption{(Color online) The fraction of the population associated
  with the Fesh\-bach resonance level of the closed channel as a function of 
  the magnetic field
  strength $B$ and the density $n$. The solid curve corresponds to the density
  characteristic for the experiment of Ref.~\cite{Regal04-2} while the
  result for the lowest density studied (the dot-dashed curve) is
  practically indistinguishable from the closed-channel admixture to
  the coupled-channels bound state wave function of the two-body
  problem. The inset enlarges the region in the vicinity of the
  zero-energy resonance. We note that, regardless of the density, the
  maximum population associated with the bosonic field operator $b_0$
  is only 8\% for $B-B_0\approx-7.2$ G. Given the virtual
  density independence of this maximum population, this result also
  coincides with the maximum closed-channel admixture to the
  highest-excited vibrational diatomic bound state of the two-channel
  two-body Hamiltonian (\ref{H2B2channel}).}
  \label{fig:occup}
\end{figure}

At the two-body level, the similarity of the results obtained from both
many-body approaches is principally due to the fact that the admixture of
the closed channel configuration to the highest excited multi-channel 
vibrational bound state is rather small for the entire range of 
experimentally relevant magnetic field strengths. 
In the many-body analysis of the boson-fermion approach this phenomenon is 
reflected in the small population associated with the bosonic field operator 
$b_0$, as shown in Figure \ref{fig:occup}. In the vicinity of the position of 
the zero-energy resonance (i.e.~for magnetic field strength differences 
$B-B_0$ from -1 G to 1 G) this population exhibits some density dependence, 
but its magnitude is always less than 4\%. In particular, for the
density relevant to the experiment of Ref.~\cite{Regal04-2} the
Fesh\-bach resonance population above $B_0$ reaches at most 0.6\%, which
is about one order of magnitude lower than the predictions of
Refs.~\cite{Stoof,Mackie05}. Further below $B_0$, it becomes 
essentially density independent and reaches a maximum of about 8\% at 7.2 G 
below the zero-energy resonance. This result agrees with the predictions 
of two-body coupled-channels calculations \cite{Julienne}. 
For even smaller values of the magnetic field strength the population 
associated with $b_0$ decreases. At the two-body level, this phenomenon stems 
from the influence of the anti-crossing between the two highest excited 
vibrational bound states (solid and dashed curves in Figure \ref{fig:E-1}). 
The domination of the entrance channel components in the states of 
weakly-bound molecules reflects their long-range nature \cite{TKTGPSJKB03}. 
%This, on the other hand, results in small overlaps with deeper bound states 
%and suppresses the de-excitation into them. Consequently, the weakly-bound
%molecules produced in the experiments with ultra-cold gases have
%comparatively long lifetimes in the environment of the gas.
%

\subsection{Pair wave function and single-particle excitation spectra}
%In the following, we study the BCS-BEC cross-over for a two-component gas of
%$^{40}$K atoms as a function of the magnetic field strength at a given
%density of $1.5\times 10^{13}$ cm$^{-3}$. 
%This density exactly recovers the Fermi energy of the trapped system in 
%Ref.~\cite{Regal04-2}. 
Several theoretical studies \cite{Keldysh,Eagles,Leggett} have predicted that 
the cross-over between the BCS phase of correlated fermion pairs and the 
BEC phase of tightly bound diatomic molecules may be realised either by 
increasing the strength of the inter-particle interactions (the route 
employed in dilute Fermi gases) or by decreasing the particle density 
(as in the case of exciton and polariton BECs). In the following, we shall 
illustrate, as a function of density and interaction strength, the different 
pairing phenomena in terms of thermal equilibrium physical quantities that 
can be derived from the mean field approach of Section \ref{sec:therm}. We 
shall show, in particular, that the BCS-BEC cross-over point (zero of the 
chemical potential) in the experiments of Ref.~\cite{Regal04-2} is shifted 
with respect to the position of the zero-energy resonance (singularity of the 
two-body scattering length) towards lower magnetic field strengths. This 
observation corroborates the conventional character of the BCS-BEC cross-over 
in $^{40}$K as opposed to the ideas of resonance superfluidity outlined in 
Refs.~\cite{Stoof,Mackie05}.

The equilibrium quantities that we shall consider involve the pair wave 
function  
\begin{equation}
  \kappa_{\mathbf{k}}=u_{\mathbf{k}}v_{\mathbf{k}}=\frac{\Sigma(\mathbf{k})}
	{2\sqrt{(\epsilon_{\mathbf{k}}-\mu)^2+ 
	    |\Sigma(\mathbf{k})|^2}} \, ,
	\label{ukvk}
\end{equation}
the fermionic distribution function
\begin{equation}
  |v_{\mathbf{k}}|^2=\frac{1}{2}
  \left[1-
    \frac{\epsilon_{\mathbf{k}}-\mu}
	 {{{\sqrt{(\epsilon_{\mathbf{k}}-\mu)^2+ |\Sigma(\mathbf{k})|^2}}}} 
	 \right]\, , 
  \label{vk}
\end{equation}
and the single-particle excitation spectrum
\begin{equation}
  E_{\mathbf{k}}=\sqrt{(\epsilon_{\mathbf{k}}-\mu)^2+
    |\Sigma(\mathbf{k})|^2}\, .
  \label{Ek}
\end{equation}

\begin{figure}[htbp]
  \centering
  \includegraphics[width=0.983\columnwidth,clip]{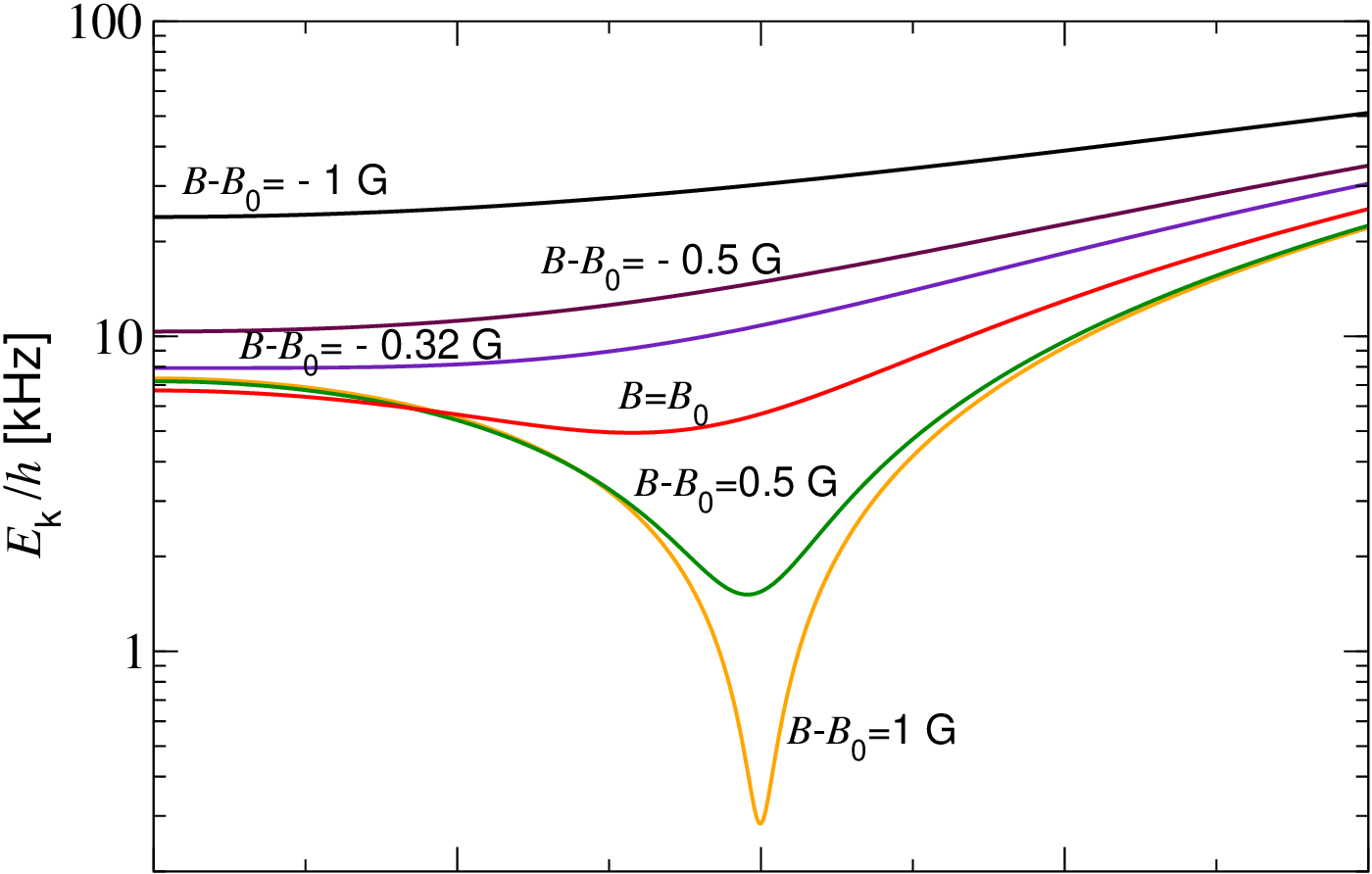}
  \includegraphics[width=\columnwidth,clip]{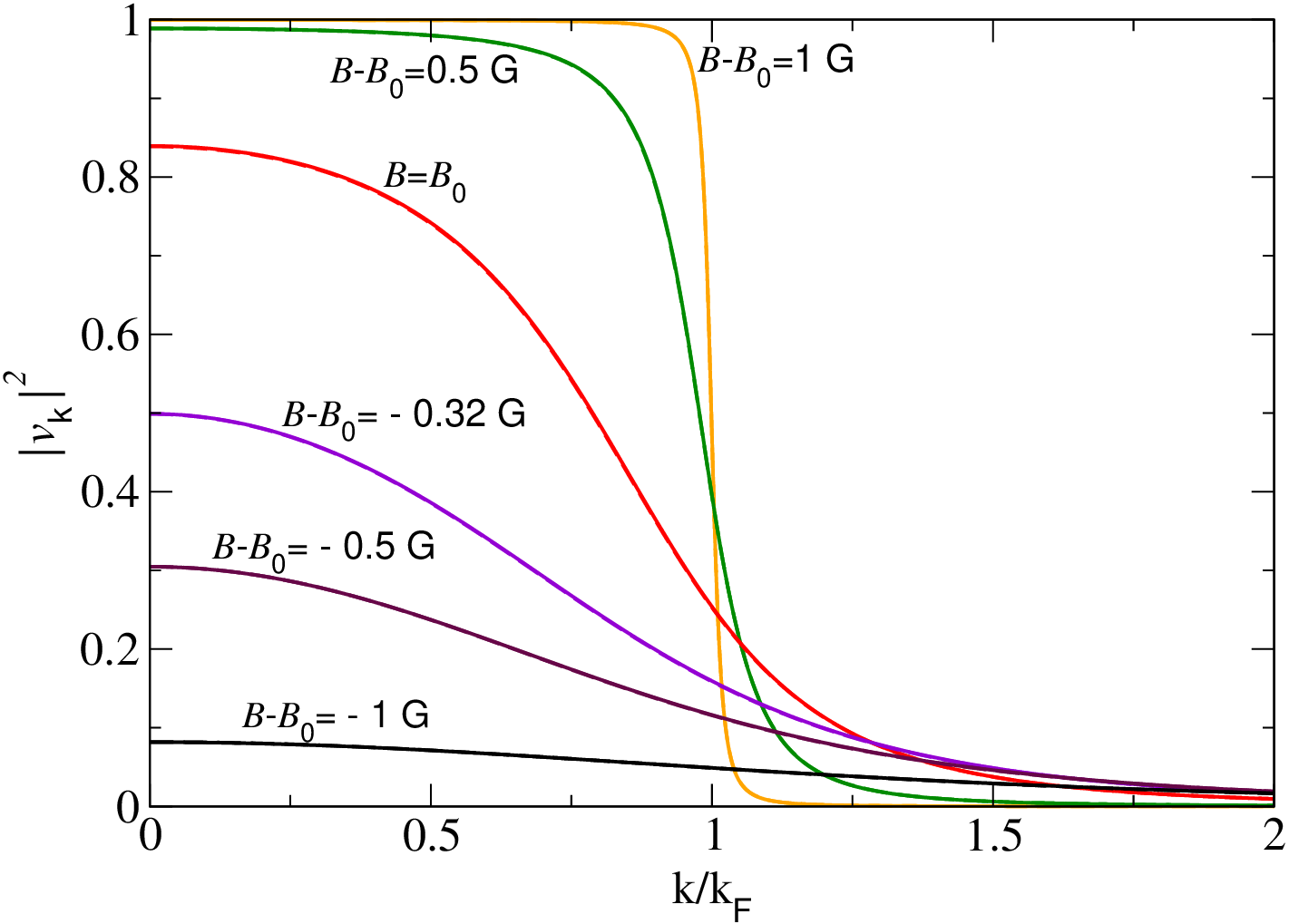}
  \caption{(Color online) The single-particle excitation spectrum (the
    upper panel), given by Eq.~(\ref{Ek}), and the fermionic momentum
    distribution function (the lower panel), determined by Eq.~(\ref{vk}), 
    for the density $n=1.5\times 10^{13}$ cm$^{-3}$ and a range of magnetic
    field strengths $B$, relevant to the experiment of
    Ref.~\cite{Regal04-2}. We note that the minimum of the excitation
    spectrum evolves from its position at $k=k_\mathrm{F}$ on the BCS side 
    to $k=0$ at $B-B_0=-0.32$ G.}
  \label{fig:spectrum} 
\end{figure}

\begin{figure}[htbp]
  \centering
  \includegraphics[width=\columnwidth,clip]{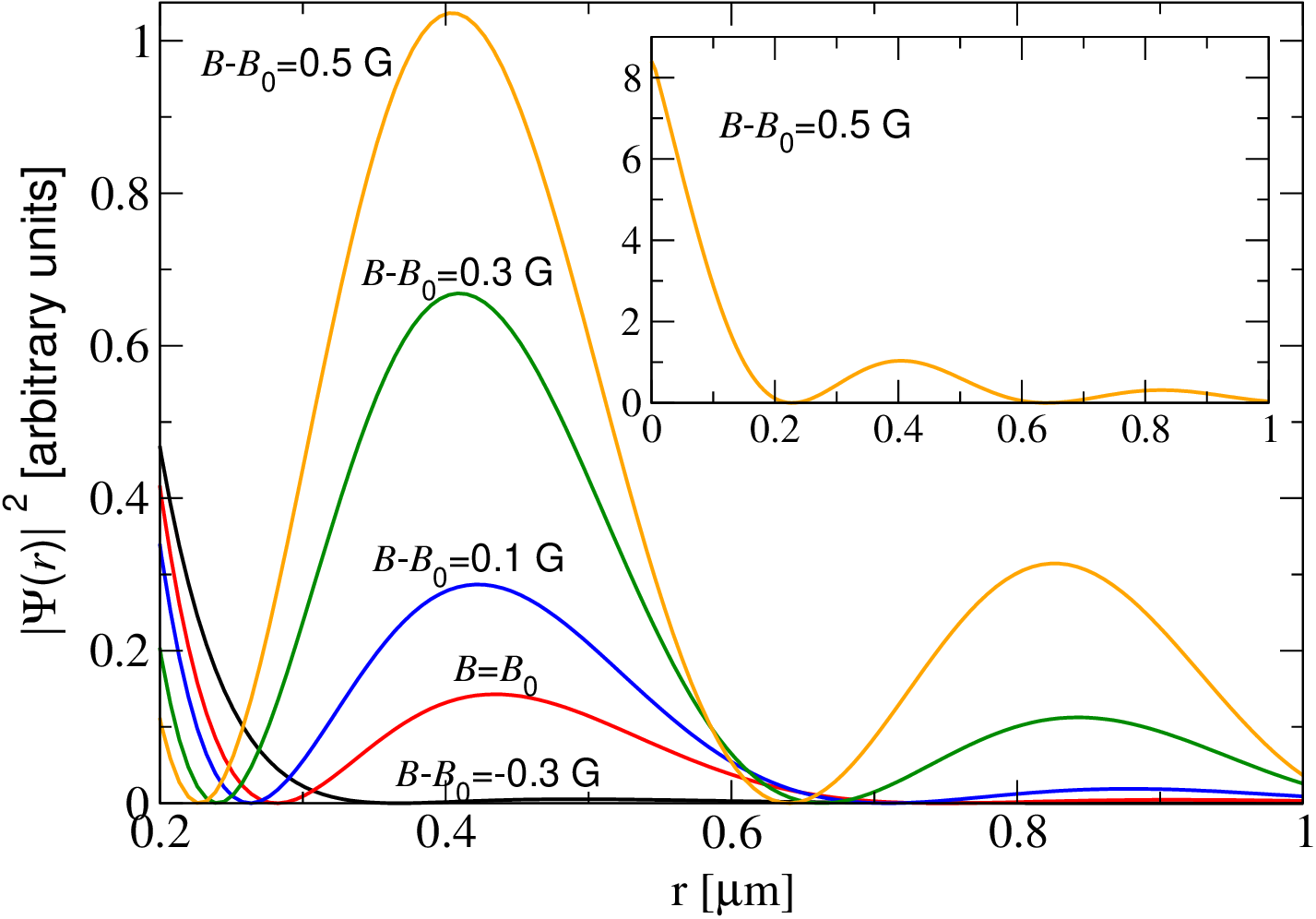}
  \caption{(Color online) The spatial form of the pair wave function
    for the density $n=1.5\times 10^{13}$ cm$^{-3}$ and a range of
    magnetic field strengths $B$. We note that for $B-B_0$ smaller than
    about -0.3 G the wave function loses its oscillatory form
    characteristic for the Cooper pairs and decays as a
    function of the inter-atomic distance. Such a virtually exponential decay
    is characteristic for diatomic bound states.}
  \label{fig:waveinr}
\end{figure}

The single-particle excitation spectra for densities of the experiment
of Ref.~\cite{Regal04-2} and different values of the magnetic field
strength (the upper panel of Figure \ref{fig:spectrum}) show
qualitatively different forms in the BEC and BCS limits of the
cross-over problem. On the BCS side the excitation spectrum has a minimum at 
non-zero $k=k_\mathrm{F}$ (the two lower curves in the upper panel of 
Fig.~\ref{fig:spectrum}), where $k_\mathrm{F}$ is the Fermi wave number. 
On the BEC side (the two upper curves in the upper panel of 
Fig.~\ref{fig:spectrum}) 
$E_\mathbf{k}$ recovers the binding energy of a molecule at its minimum at 
$k=0$. For finite wave numbers $k$ the spectrum has a quadratic form 
associated with the dispersion relation of the kinetic energy of the 
molecules. We note that at a finite density the position of the zero-energy 
resonance $B_0$ is not a characteristic field strength with respect 
to the cross-over problem. In fact, at $B=B_0$ the spectrum shows BCS-like 
features due to its minimum at non-zero $k$. The value of the magnetic field 
strength where the minimum of the spectrum changes from non-zero to zero 
$k$ depends on the density. In particular, for the density of the experiment of
Ref.~\cite{Regal04-2} this characteristic magnetic field strength is located 
at approximately 0.32 G below $B_0$. The fermionic momentum distribution 
function 
$|v_{\mathbf{k}}|^2$, as given by Eq.~(\ref{vk}), is shown in the lower
panel of Fig.~\ref{fig:spectrum}. It smoothly evolves, as a function of the 
magnetic field strength, from the step-like form of a weakly-interacting Fermi 
gas (the BCS phase) towards a flat distribution
characteristic for the macroscopic occupation of the lowest energy mode in 
a condensed Bose gas (the BEC phase). The qualitative change of 
pairing phenomena between the BCS and BEC limits is most intuitively reflected 
in the spatial form of the pair wave function. Its oscillatory behaviour on 
the BCS side is characteristic for unbound correlated  pairs, while its 
exponential 
decay in the BEC limit recovers the long range asymptotic form of the 
molecular bound state wave function (see Fig.~\ref{fig:waveinr}). 

\begin{figure}[htbp]
  \centering
  \includegraphics[width=\columnwidth,clip]{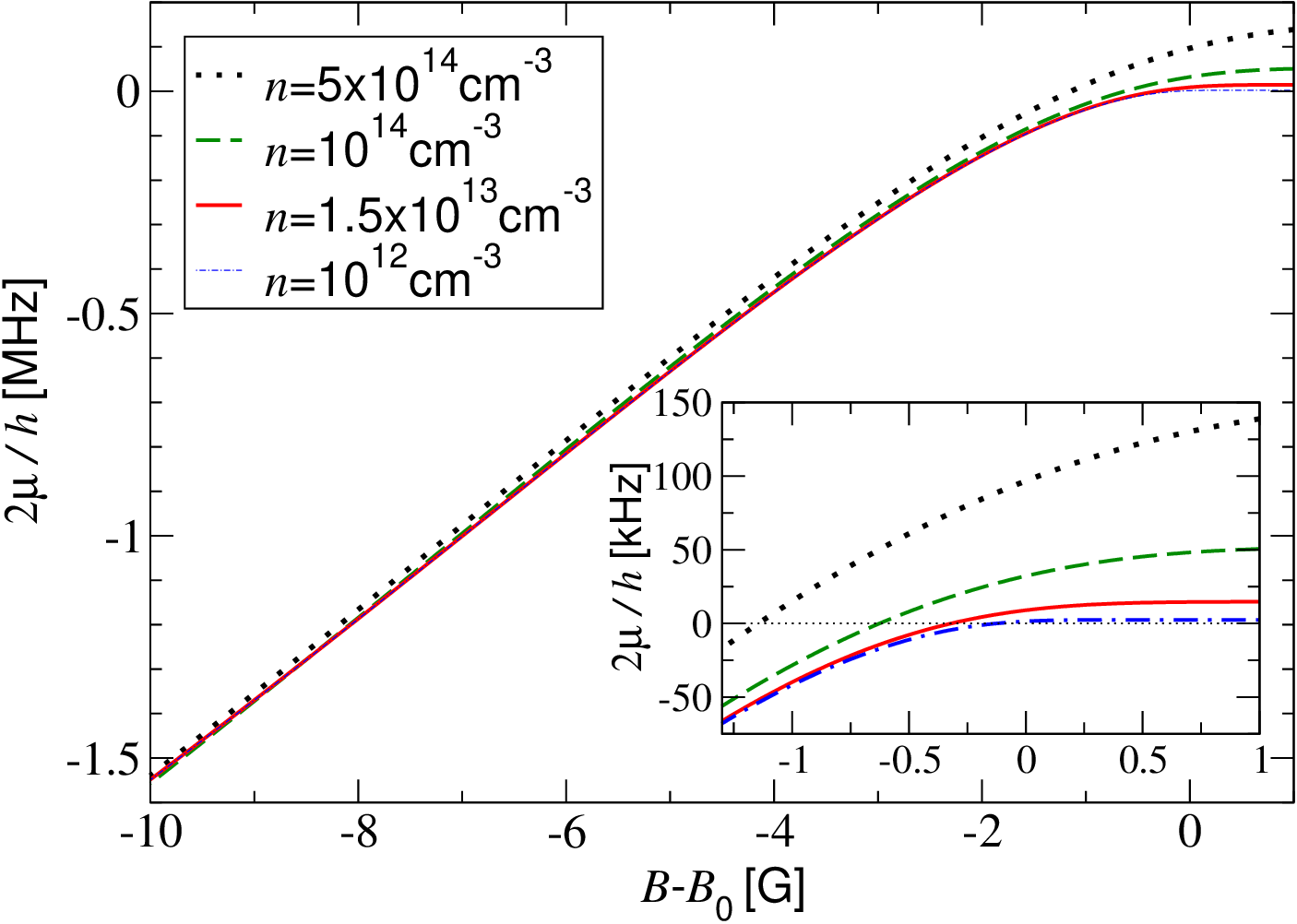}
  \caption{(Color online) The chemical potential $\mu$ as a function
    of the magnetic field strength $B$ for different densities
    $n$. Away from $B_0$, in the BEC limit, $2\mu$ approaches
    the energy of the highest excited vibrational molecular
    bound state $E_\mathrm{b}$, as depicted in Figure
    \ref{fig:40K}. We note that for the density $n=1.5\times
    10^{13}$ cm$^{-3}$ of the experiment \cite{Regal04-2} the
    chemical potential crosses zero at $B-B_0=-0.32$ G. The
    inset enlarges the region in the vicinity of the
    zero-energy resonance.}
  \label{fig:chem}
\end{figure}

In the case of contact interactions, which are determined by a single 
parameter involving the scattering length, Leggett has shown
\cite{Leggett} that the qualitative changes in the form of the pair
wave function and the excitation spectrum occur when the chemical
potential $\mu$ crosses zero. The chemical potential is positive on
the BCS side and negative on the BEC side of the cross-over. In the 
case of spatially extended or non-local potentials (such as separable 
interactions) the qualitative change in the form of the pair wave function and
the excitation spectra does not, in general, coincide with the vanishing
of the chemical potential.
%the definition of the boundary between the BCS and BEC 
%regimes generally depends on the density.
Our results indicate, however, 
that in a dilute gas of $^{40}$K atoms at typical experimentally accessible 
densities this boundary practically coincides with Leggett's universal 
boundary at $\mu=0$. It is therefore instructive to study the chemical
potential as a function of the magnetic field strength and as a function of
the density (see Figure \ref{fig:chem}).
Far below $B_0$ the chemical potential is negative, independent of the
density and its limiting value recovers one half of the energy of the highest
excited vibrational molecular bound state. The magnetic field strength
where the chemical potential crosses zero (the cross-over point) depends on 
the density. Given the density $n=1.5\times 10^{13}$\,cm$^{-3}$, 
characteristic for the experiments of Ref.~\cite{Regal04-2}, the cross-over 
point is located at 0.32 G below the position of the zero-energy resonance.

\begin{figure}[htbp]
  \centering
  \includegraphics[width=\columnwidth,clip]{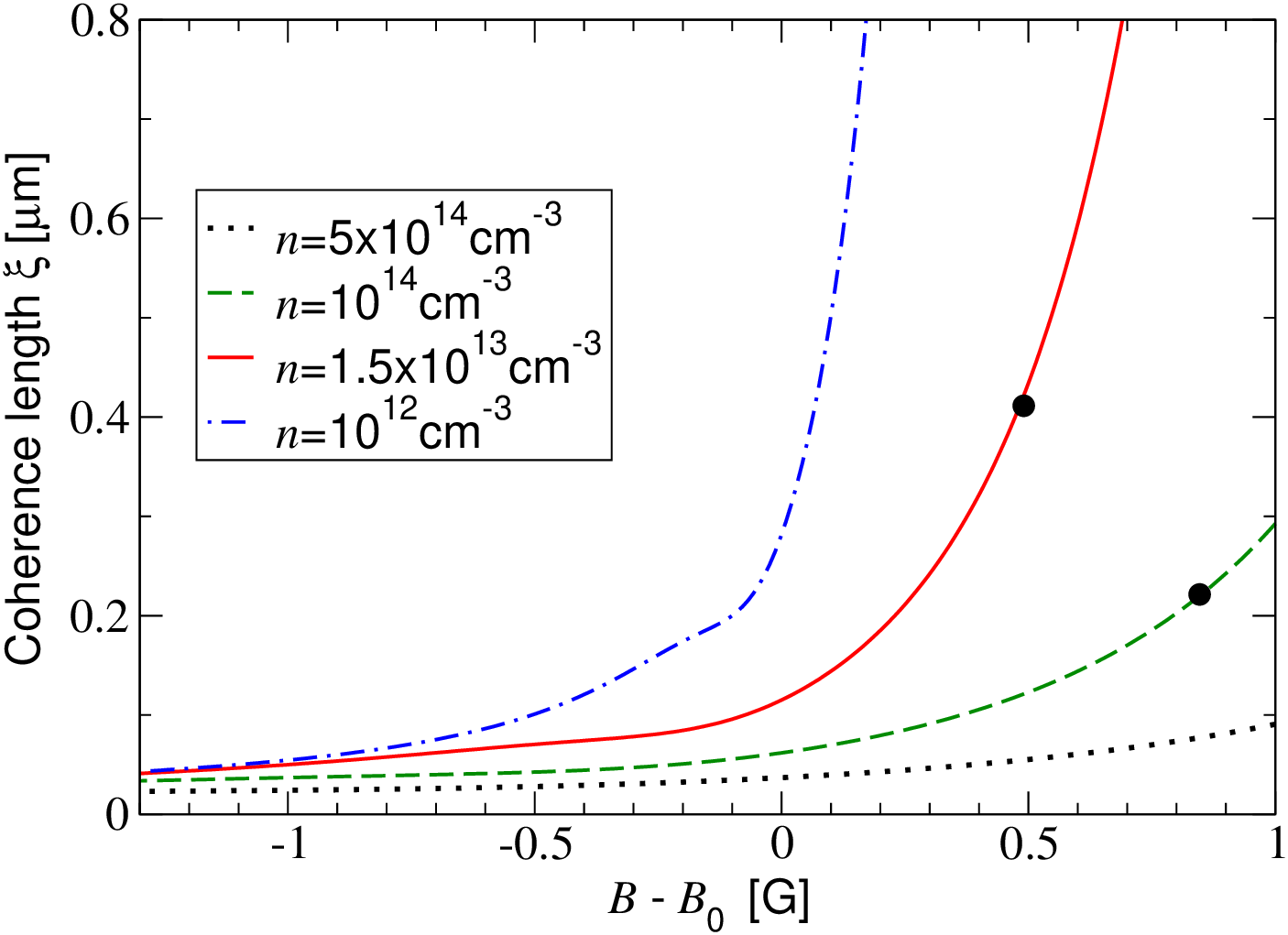}
  \caption{(Color online) The coherence length $\xi$, characterising
    the size of the fermion pairs, as a function of the magnetic field 
    strength $B$ for different densities $n$. The circles indicate the 
    values of the magnetic field strength at which the size of the pairs
    equals the mean inter-atomic spacing. We note that for the density
    $n=1.5\times10^{13}$ cm$^{-3}$ of the experiment of
    Ref.~\cite{Regal04-2} (solid curve) the size of the pairs 
    equals the mean inter-atomic spacing at $B\approx B_0+0.5$ G.}
  \label{fig:pairsize}
\end{figure}

To interpret the results of the pairwise projection technique employed 
in Ref.~\cite{Regal04-2}, we have also analysed the magnetic 
field and density dependences of the coherence length $\xi$. This 
typical length scale characterises the size of the fermionic pairs and is 
given by:
\begin{equation}
  \xi^2=\frac{\int d\mathbf{r}\, r^2\, [\kappa(r)]^2}{\int  d\mathbf{r}\, 
    [\kappa(r)]^2} \, .
  \label{xi}
\end{equation}
Figure \ref{fig:pairsize} shows the coherence length of the fermion
pairs and the magnetic field strength for which the size of the pairs
equals the mean inter-atomic spacing. For the characteristic density of the 
experiment of Ref.~\cite{Regal04-2} the size of the fermion pairs 
equals the mean inter-atomic spacing at a magnetic field strength of about 
0.5 G above the position of the zero-energy resonance. This coincides 
with the boundary value of the magnetic field strength at which 
condensation phenomena were observed via the pairwise 
projection technique employed in the experiments of 
Ref.~\cite{Regal04-2}. 
%In Subsection \ref{subsec:pairwiseproj}
%we shall address the question of whether probing the fermionic pair 
%condensates by pairwise projection onto molecules can be successfully 
%used in the BCS regime where fermionic pairs spatially overlap.
%

\subsection{Pairwise projection of fermionic condensates onto molecules}
\label{subsec:pairwiseproj}
In the experiments of Ref.~\cite{Regal04-2} a new method of probing the
paired fermionic component of the gas was used. This technique involved 
a fast ramp of the magnetic field strength across the 202 G resonance of
$^{40}$K. Ideally, this rapid process projects the centre-of-mass
momentum distribution of the correlated fermionic pairs on the high field side 
of $B_0$ onto the centre-of-mass momentum distribution of weakly bound 
molecules on the low-field side of the zero-energy resonance \cite{KGKB04}. 
Such a trapped fermionic condensate of correlated pairs is characterised 
by a narrow centre-of-mass momentum spread, which was indeed observed 
as a distinct peak in the measured molecular momentum distribution
\cite{Regal04-2}. In the practical implementation of this procedure
reported in Ref.~\cite{Regal04-2}, the magnetic field strength was swept to
about 10 G below the position of the zero-energy resonance.

\begin{figure}[htbp]
  \centering
  \includegraphics[width=\columnwidth,clip]{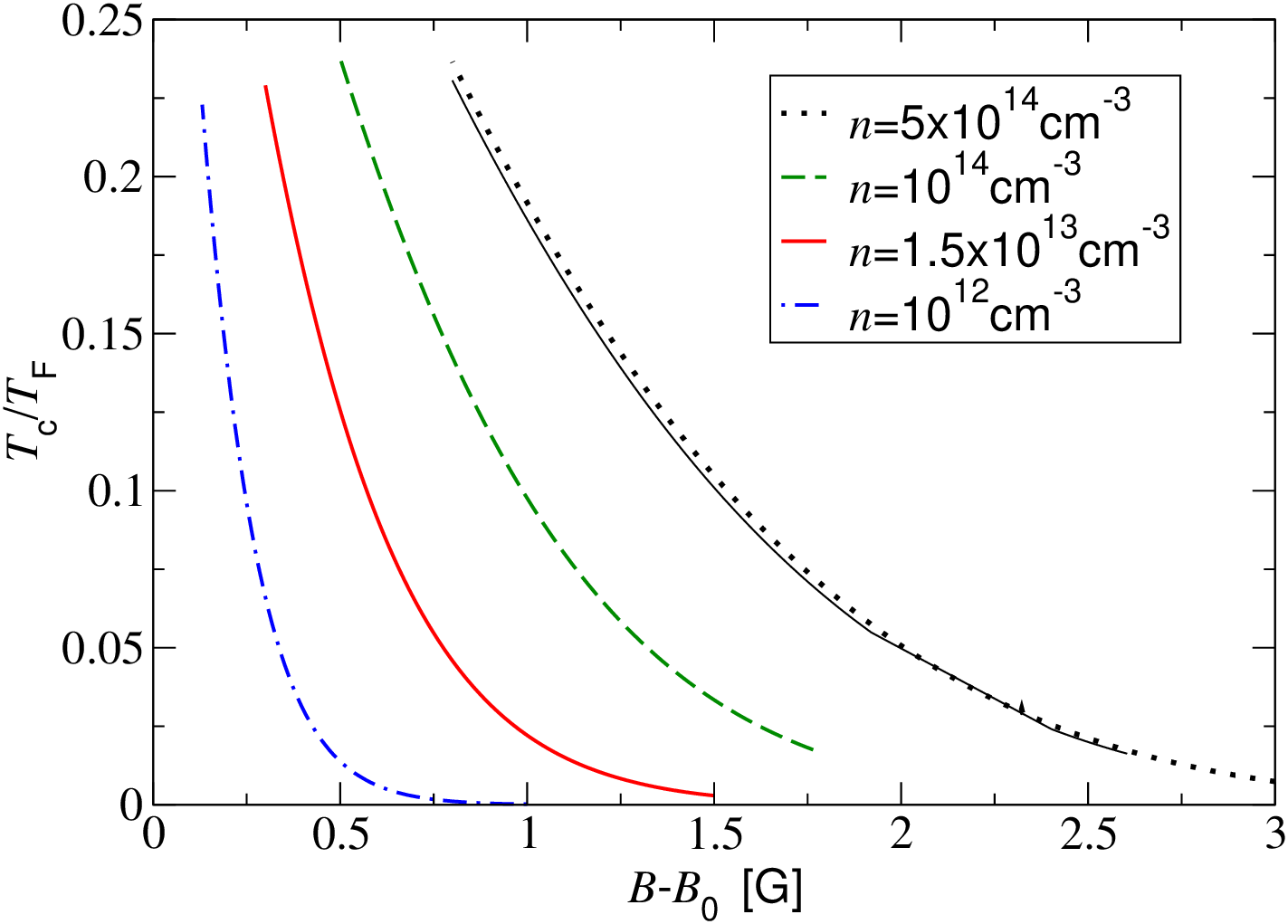}
  \caption{(Color online) The critical temperature $T_\mathrm{c}$ for
    the BCS transition as a function of the magnetic field strength
    $B$ for a range of densities $n$. The density of $n=1.5\times
    10^{13}$ cm$^{-3}$ corresponds to the Fermi temperature of 
    $T_\mathrm{F}=0.35\, \mu$K quoted in Ref.~\cite{Regal04-2}. 
    We note the slight differences between the predictions of the 
    boson-fermion (thin solid curve) and standard fermionic (dotted curve) 
    for the highest density of $n=5\times 10^{14}$\,cm$^{-3}$. This
    high-density critical temperature curve reflects the only calculation 
    for which we were able to identify any visible differences between
    the two approaches.
  %solid black curve is 1 channel and dotted is BF
  }
  \label{fig:Tc}
\end{figure}
\begin{figure}[htbp]
  \begin{center}
\includegraphics[width=\columnwidth,clip]{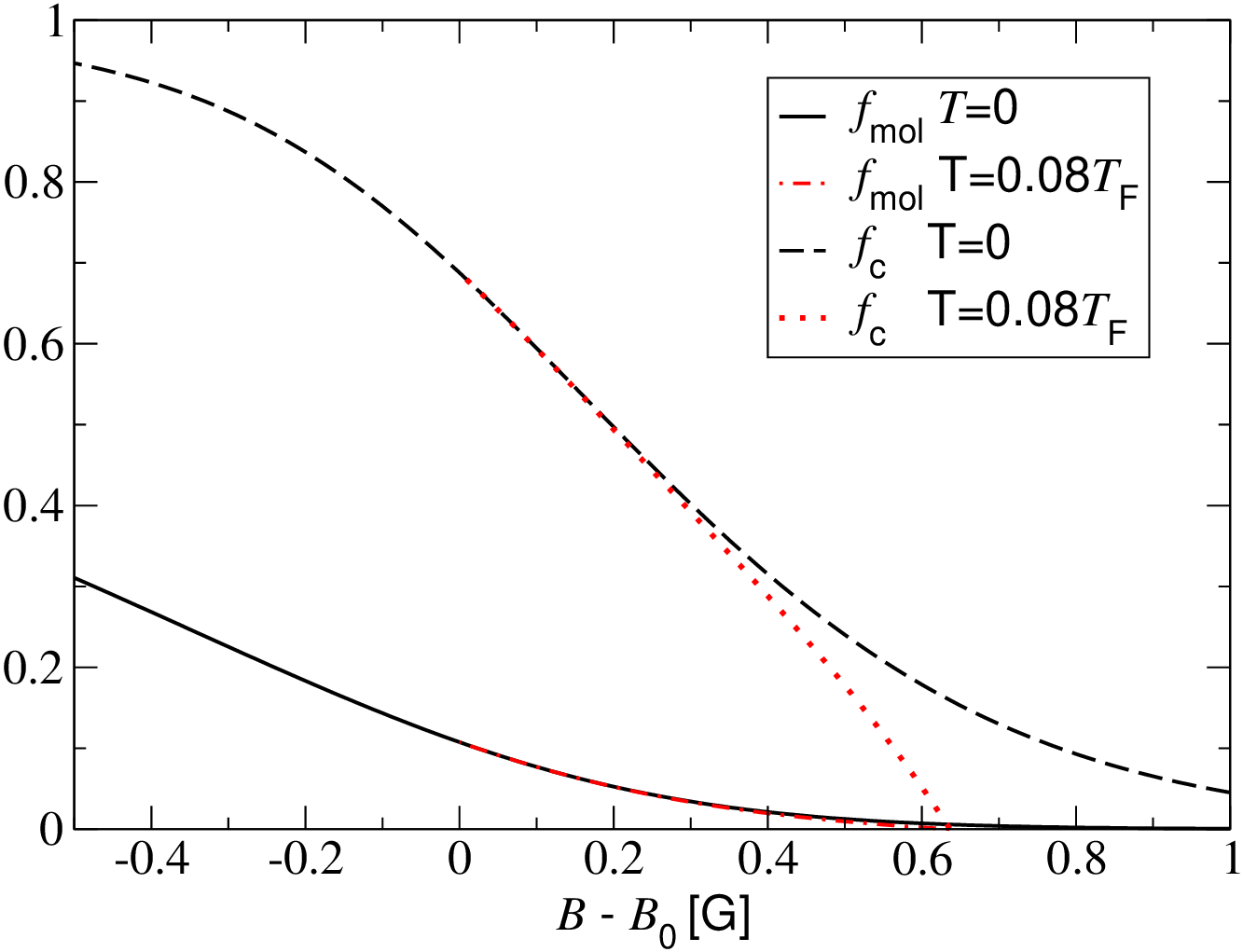}
\end{center} 
 \caption{(Color online) The fermionic condensate fraction
 $f_\mathrm{c}$ and its overlap $f_{\mathrm{mol}}$ with the target
 molecular bound state at 10 G below the zero-energy resonance for
 $T=0$ (dashed and solid curves, respectively) and for
 $T=0.08T_\mathrm{F}$ (dotted and dot-dashed curves, respectively). 
 The density is $n=1.5\times 10^{13}$ cm$^{-3}$
 corresponding to the Fermi temperature of 
 $T_\mathrm{F}=0.35 \mu$K quoted in Ref.~\cite{Regal04-2}.}
  \label{fig:mol}
\end{figure}

In order to gain further insight into the range of applicability of the
experimental pairwise projection technique, we have analysed the
critical temperature $T_{\mathrm{c}}$ for the BCS transition
(depicted in Figure \ref{fig:Tc}) as well as
the fermionic condensate fraction $f_{\mathrm{c}}$, i.e. the density
of condensed pairs divided by one half of the total atomic density.
This quantity $f_\mathrm{c}$ is thus given by
\begin{equation}
  f_\mathrm{c}=\frac{|b_0|^2+
    \int d\mathbf{p} \, |\kappa_p|^2 }{n/2} \, .
  \label{nc}
\end{equation}
One of the measurable quantities in the pairwise projection technique 
is the density of condensed molecules observed after the fast sweep 
of the magnetic field strength below the position of the zero-energy
resonance. For an idealised, i.e. instantaneous, sweep, this condensate 
density would be determined just by the projection of the initial pair 
wave function onto the wave function of the highest excited vibrational 
bound state $\phi_\mathrm{b}(B_\mathrm{proj})$ [i.e.~an exact molecular
bound state of the two-channel two-body Hamiltonian (\ref{H2B2channel})] 
at the final magnetic field strength $B_\mathrm{proj}$. This Fesh\-bach 
molecular state becomes populated by the sweep. In the case of 
Ref.~\cite{Regal04-2} the final magnetic field strength was 
$B_\mathrm{proj}=B_0-10$ G. We have normalised the density of 
condensed molecules by a half of the atomic density $n$, in analogy to the 
definition of the fermionic condensate fraction of Eq.~(\ref{nc}). This yields:
\begin{equation}
  f_\mathrm{mol}=\frac{\left|b_0/\mathcal{N}_\mathrm{b}
    +\int d\mathbf{p} \ \kappa_p \,
    \phi_\mathrm{b}^\mathrm{bg}(p,B_\mathrm{proj})\right|^2}{n/2} \, ,
  \label{nmc}
\end{equation}
where $\phi_\mathrm{b}^\mathrm{bg}$ is the entrance-channel component
of the two-channel Fesh\-bach molecular bound state given by Eq.~(\ref{phib}) 
and $\mathcal{N}_\mathrm{b}$ is the bound-state normalisation constant defined 
in Eq.~(\ref{twochannelnormalisation}).
The fermionic condensate fraction $f_\mathrm{c}$ and its overlap
$f_\mathrm{mol}$ with the bound state wave function $
\phi_\mathrm{b}(B_\mathrm{proj})$ are shown in Fig.~\ref{fig:mol} (see
also \cite{Avdeenkov05}).

As shown in Fig.~\ref{fig:mol}, at the lowest experimentally
reported \cite{Regal04-2} temperature of 0.08 $T_\mathrm{F}$ the
fermionic condensate should be present up to about 0.6 G above the
zero-energy resonance. For magnetic field strengths larger
than about $B_0+0.5$ G, however, the overlap of the fermionic 
pair wave function with the target molecular wave function
%it is projected into 
is negligible. Consequently, the paired fraction at
higher magnetic fields is not detected by the pairwise projection
technique, at least, under the idealising assumption 
that the magnetic field sweep is infinitely fast. The magnetic field 
strength at which the overlap of the paired fraction
with the target molecular state vanishes also coincides with the
field strength at which the extent of the pairs equals the mean 
inter-atomic spacing of the gas.

\section{Conclusions}
\label{sec:concl}

We have shown in this paper that the parameters of the general 
standard fermionic and boson-fermion many-body Hamiltonians, 
commonly used in studies of the BCS-BEC cross-over problem 
in dilute atomic gases, can be determined in an unambiguous manner by
applying the Hamiltonians to a single pair of atoms and determining the 
low-energy binary bound state and scattering properties. In particular, 
we have described a convenient two-channel approach to the resonance enhanced 
two-body scattering, involving separable inter-atomic interactions. In a 
possible extension of this approach to the many-body description of a gas 
the five parameters of the boson-fermion Hamiltonian are all determined 
through five physical parameters of the resonance enhanced binary scattering, 
all of which can be deduced from measurements of binary scattering properties.

We have studied in detail the 202 G resonance in a mixture of
$(f=9/2,m_f=-9/2)$ and $(f=9/2,m_f=-7/2)$ Zeeman states of $^{40}$K 
fermions and shown that the maximum closed-channel admixture in the
highest excited vibrational bound state reaches only about 8\% at 7.2
G below the position of the zero-energy resonance. This small 
admixture is due to an avoided crossing between the energies of the 
Fesh\-bach resonance level and a bound state of the background 
scattering potential. This points to the necessity of including 
the highest excited vibrational state of the background-scattering potential, 
in addition to the background scattering length, in the description of binary 
scattering in the vicinity of the zero-energy resonance relevant to the 
BCS-BEC cross-over. Including this second parameter characterising the 
entrance channel scattering also significantly extends the range of 
applicability of effective single channel approaches, a fact that
often appears not to be appreciated. In fact, for the $^{40}$K system 
studied in our paper we 
conclude that the energies of the highest-excited molecular bound state 
given by the effective single-channel approach are virtually 
indistinguishable from the results of a more elaborate two-channel 
analysis over the whole range of magnetic fields relevant to the 
experiment of Ref.~\cite{Regal04-2}. The universal regime in the vicinity 
of the zero-energy resonance, however, where the binding energy is a 
function of the scattering length only, covers only a rather small fraction 
of the BCS-BEC regime in the case of $^{40}$K.

These observations suggest that the effective single-channel approach
with only two parameters should be sufficient also in studying the
many-body BCS-BEC cross-over problem. To verify this, we have
performed the mean-field analysis of the cross-over regime using the
standard fermionic and the commonly used boson-fermion
Hamiltonians for a wide range of atomic densities and magnetic
fields. As anticipated, the results of our comparative studies remain
virtually indistinguishable for all the many-body observables common
to both types of Hamiltonians. The population
associated with the bosonic field of the boson-fermion Hamiltonian 
remains, for the whole range of experimentally relevant densities, very
close to the closed channel populations of the Fesh\-bach molecule
predicted by the two-body two-channel approach. In particular,
in the regime where it reaches its maximum value of only about 8\% (at
7.2 G below $B_0$) this quantity is already virtually density
independent. For $B\geq B_0$ and the density of the experiment of
Ref.~\cite{Regal04-2} it remains below 0.6\%, in stark contrast
to the results of Refs.~\cite{Stoof,Mackie05}. We therefore conclude
that for the $^{40}$K system the standard fermionic many-body
approach should be sufficient to study the whole BCS-BEC cross-over
regime, as long as it properly accounts for the scattering 
length and the energy of the highest-excited vibrational bound state of the 
background-scattering potential. 

Adjusting the parameters of the many-body Hamiltonian on the basis of the
two-body considerations puts the proceeding analysis of the BCS-BEC
cross-over problem of $^{40}$K on firm ground. Adopting a different procedure
may lead to unphysical conclusions. For example,
one may reach a conclusion that below the zero-energy resonance there is a 
large population associated with the quantum field describing structureless 
bosons, as it appears in the boson-fermion Hamiltonian. One can also 
mistakenly associate this population with the number of the actually produced 
weakly-bound molecules. Such conclusions may result purely from
ignoring the dominant influence of the background scattering potential and its
highest excited vibrational bound state of energy $E_{-1}$. We have demonstrated 
that the background-scattering potential and its highest-excited vibrational 
bound state are crucial to a proper description of both 
two- and many-body physics outside the small universal regime in the
vicinity of the zero-energy resonance.

We would also like to emphasise
that the position $B_0$ of the zero-energy resonance, where the weakly bound
molecular state emerges, is, due to strong inter-channel coupling,
very different from the magnetic field strength $B_\mathrm{res}$ where the 
energy of the Fesh\-bach resonance level $E_{\mathrm{res}}$ crosses the
dissociation threshold of the entrance channel. In fact, for the
$^{40}$K resonance at $B_0=202.1$ G, $E_{\mathrm{res}}(B_0)$ is as
large as $h\times$21.8 MHz and negative. We note that the analyses of
Refs.~\cite{Stoof,Mackie05} do not capture this important physical
fact as they identify $B_\mathrm{res}$ with $B_0$.

Finally, we have analysed thermodynamic quantities associated with 
the BCS-BEC cross-over in
$^{40}$K. We found that for densities of the experiments of 
Ref.~\cite{Regal04-2}
the cross-over point, as characterised by a qualitative change in the form
of the pair wave function from oscillatory to decaying and in the form of the
single-particle excitation spectrum, takes place about 0.3 G below
the position of the zero-energy resonance. We have also studied the pairwise
projection technique employed in the experiment of
Ref.~\cite{Regal04-2} by calculating the overlap of the fermionic condensate
wave function with the target molecular state at about 10 G below
the zero-energy resonance. We found that when the fermionic pairs start to
spatially overlap, which in the case studied takes place about 0.5
G above the position of the zero-energy resonance, the sensitivity
of the pairwise projection technique decreases. This result coincides
with the reported measurements of the condensed fraction of fermionic pairs 
extending up to about 0.5 G above the position of the zero-energy resonance.

\section{Acknowledgements}
We are grateful to Paul Julienne, Peter Littlewood, Francesca Marchetti and 
Ben Simons for stimulating discussions. This research has been 
supported by Gonville and Caius College Cambridge (M.H.S.), the UK EPSRC 
(K.B.) and the Royal Society (K.G., T.K., and K.B.).


\begin{thebibliography}{99}

\bibitem{Keldysh} 
  L.V. Keldysh and You.V. Kopaev, \emph{Fiz. Tverd. Tela} (Leningrad)
  \textbf{6}, 2791 (1964) [\emph{Sov. Phys. Solid State} \textbf{6},
    2219 (1965)]; 
  L.V. Keldysh and A.N. Kozlov, \emph{Zh. \'Eksp. Teor. Fiz.}
  \textbf{54}, 978 (1968) [\emph{Sov. Phys. JETP} \textbf{27}, 521
    (1968)].

\bibitem{Eagles}
  D.M Eagles, Phys. Rev.  \textbf{186}, 456 (1969).
  
\bibitem{Leggett} A.J. Leggett, in \emph{Modern Trends in the Theory of
  Condensed Matter}, edited by A. Pekalski and R. Przystawa
  (Springer-Verlag, Berlin, 1980).
  
\bibitem{Noz-SchR} P. Nozi\'eres and S. Schmitt-Rink,
  %Bose Condensation in an Attractive Fermion Gas: From Weak to Strong
  %Coupling Superconductivity,
  J.\ Low Temp.\ Phys. \textbf{59}, 195 (1985).
 
\bibitem{Rand} M. Randeria, in \emph{Bose-Einstein Condensation}, 
  edited by A. Griffin, D.W. Snoke, and S. Stringari 
  (Cambridge University Press, Cambridge, 1995).

\bibitem{Greiner03}
  %3 November 2003  K2 BEC
  M. Greiner, C.A. Regal, and D.S. Jin, Nature (London) \textbf{426},
  537 (2003).

\bibitem{Grimm-mol}
  %4 November 2003 BEC Li2 molecules
  S. Jochim, M. Bartenstein, A. Altmeyer, G. Hendl, 
  C. Chin, J. Hecker Denschlag, and R. Grimm,
  %Bose-Einstein Condensation of Molecules,
  Science \textbf{302}, 2102 (2003).

\bibitem{Ket-mol}
  %27 November 2003 BEC Li2 molecules
  M.W. Zwierlein, C.A. Stan, C.H. Schunck,
  S.M.F. Raupach, S. Gupta, Z. Hadzibabic, and W. Ketterle,
  %Observation of Bose-Einstein condensation of molecules,
  Phys. Rev. Lett. \textbf{91}, 250401 (2003).
  
\bibitem{Regal04-2}
  %13 January K fermionic condensate
  C.A. Regal, M. Greiner, and D.S. Jin, Phys. Rev. Lett. \textbf{92},
  040403 (2004).

\bibitem{Zwierlein04}
  %1 March 2004 Li fermionic condensate
  M.W. Zwierlein, C.A. Stan, C.H. Schunck, S.M.F. Raupach,
  A.J. Kerman, and W. Ketterle, Phys. Rev. Lett. \textbf{92}, 120403
  (2004).

\bibitem{Bartenstein04}
  %8 January 2004  Li cross-over
  M. Bartenstein, A. Altmeyer, S. Riedl, S. Jochim, C. Chin, J. Hecker
  Denschlag, and R. Grimm, Phys. Rev. Lett. \textbf{92}, 120401 (2004).

\bibitem{Bourdel04}
  %Li cross-over
  T. Bourdel, L. Khaykovich, J Cubizolles, J. Zhang, F. Chevy,
  M. Teichmann, L. Tarruell, S.J.J.M.F. Kokkelmans, and C. Salomon,
  Phys. Rev. Lett. \textbf{93}, 050401 (2004).

\bibitem{Kinast04}
  J. Kinast, S.L. Hemmer, M.E. Gehm, A. Turlapov, and J.E. Thomas,
  Phys. Rev. Lett. \textbf{92}, 150402 (2004).

\bibitem{Chin04}
  C. Chin, M. Bartenstein A. Altmeyer, S. Riedl, S. Jochim, J. Hecker
  Denschlag, and R. Grimm, Science \textbf{305}, 1128 (2004).

\bibitem{Greiner04}
  M. Greiner, C.A. Regal, and D.S. Jin, Phys. Rev. Lett. 
  \textbf{94}, 070403 (2005).

\bibitem{Zwierlein05}
  M.W. Zwierlein, C.H. Schunck, C.A. Stan, S.M.F. Raupach, and
  W. Ketterle, Phys. Rev. Lett. \textbf{94}, 180401 (2005).

\bibitem{Holland}
  M. Holland, S.J.J.M.F. Kokkelmans, M.L. Chiofalo, and R. Walser, 
  %Resonance Superfluidity in a Quantum Degenerate Fermi Gas,
  Phys. Rev. Lett. \textbf{87}, 120406 (2001).
  
\bibitem{Timm}
  E. Timmermans, K. Furuya, P.W. Milonni, and A.K. Kerman,
  %Prospect of creating a composite Fermi-Bose superfluid,
  Phys.\ Lett.\ A \textbf{285}, 228 (2001).

\bibitem{Ranninger85}
  J. Ranninger and S. Robaszkiewcz, 
  Physica B \textbf{135}, 468 (1985).

\bibitem{Friedberg89}
  R. Friedberg and T.D. Lee, 
  Phys. Rev. B \textbf{40}, 6745 (1989).

\bibitem{KGB03} 
  T.~K\"ohler, T.~Gasenzer, and K.~Burnett, Phys.~Rev.~A
  \textbf{67}, 013601 (2003).

\bibitem{TKTGPSJKB03}
  T.~K\"ohler, T.~Gasenzer, P.S.~Julienne, and K.~Burnett,
  Phys.~Rev.~Lett.~\textbf{91}, 230401 (2003).

\bibitem{Schrieffer}
  J. R. Schrieffer, \emph{Theory of Superconductivity} 
  (Perseus Publishing, Cambridge, Massachusets, 1999).

\bibitem{GKGTJ03} 
  K.~G\'oral, T.~K\"ohler, S.A.~Gardiner, E.~Tiesinga,
  and P.S.~Julienne, J. Phys. B \textbf{37}, 3457 (2004).
  
\bibitem{Strecker03}
  %21 July 2003 Li Molecules
  K.E. Strecker, G.B. Partridge, and R.G. Hulet, 
  Phys. Rev. Lett.~\textbf{91}, 080406 (2003).

\bibitem{KGG04}
  T.~K\"ohler, K.~G\'oral, and T.~Gasenzer, 
  Phys. Rev. A \textbf{70}, 023613 (2004).

\bibitem{Strinati}
  S. Simonucci, P. Pieri, and G.C. Strinati, 
  Europhys. Lett. \textbf{69}, 713 (2005).
  
\bibitem{Julienne}
  P.S. Julienne, private communication.

\bibitem{Littlewood04}
  P.B. Littlewood, P.R. Eastham, J.M.J. Keeling, F.M. Marchetti, B.D. Simons, 
  and M.H. Szyma{\' n}ska, J. Phys.: Condens. Matter \textbf{16}, 3597 (2004).

\bibitem{Eastham01}
  P. R. Eastham and P. B. Littlewood, 
  Phys. Rev. B \textbf{64}, 235101 (2001).
  
\bibitem{Keeling04}
  J. Keeling, P.R. Eastham, M.H. Szyma{\'n}ska, and P.B. Littlewood,
  Phys. Rev. Lett. \textbf{93}, 226403 (2004)

\bibitem{Stoof}
  G.M. Falco and H.T.C. Stoof, 
  Phys.~Rev.~Lett.~\textbf{92}, 130401 (2004).

\bibitem{Mackie05}
  M. Mackie and J. Piilo, Phys.~Rev.~Lett.~\textbf{94}, 060403 (2005).

\bibitem{Gao98}
  Given the background scattering length $a_\mathrm{bg}$, the energy $E_{-1}$ 
  is determined solely by the van der Waals dispersion coefficient $C_6$
  characterising the (non-retarded) asymptotic behaviour 
  $V_\mathrm{bg}(r)\underset{r\to\infty}{\sim}-C_6/r^6$ of the exact
  microscopic background scattering potential at large inter-atomic distances 
  $r$ [cf.~B.~Gao, Phys. Rev. A \textbf{58},
  4222 (1998)].

\bibitem{Loftus02}
  T. Loftus, C.A. Regal, C. Ticknor, J.L. Bohn, and
  D.S. Jin, Phys. Rev. Lett. \textbf{88}, 173201 (2002).

\bibitem{Julienne89} 
  P.S.~Julienne and F.H.~Mies, 
  J.~Opt.~Soc.~Am.~B \textbf{6}, 2257 (1989).

\bibitem{Gribakin93}
  G.F.~Gribakin and V.V.~Flambaum, 
  Phys.~Rev.~A~\textbf{48}, 546 (1993).

\bibitem{Derevianko99}
  A. Derevianko , W.R. Johnson, M.S. Safronova , and J.F. Babb,
  Phys. Rev. Lett. \textbf{82}, 3589 (1999).

\bibitem{Szymanska04}
  M.H. Szyma{\'n}ska, K.~G\'oral, and T.~K\"ohler (unpublished).

\bibitem{Fricke96}
  J.~Fricke, Ann.~Phys.~(N.Y.) \textbf{252}, 479 (1996).

\bibitem{KB02}
  T.~K\"ohler and K.~Burnett, 
  Phys.~Rev.~A \textbf{65}, 033601 (2002).

\bibitem{meera}
  M.M. Parish, B. Mihaila, B.D. Simons, and P.B. Littlewood, 
  Phys. Rev. Lett. \textbf{94}, 240402 (2005).

%\bibitem{Kheruntsyan00}
%K.V. Kheruntsyan and P.D. Drummond, Phys. Rev. A \textbf{61}, 063816
%(2000) suggest the application of the boson-fermion Hamiltonian to the
%case of Fermi gases and analyse its two-body bound states as well as
%its many-body energy bounds.
  
\bibitem{Ohashi02}
  Y. Ohashi and A Griffin, 
  Phys.~Rev.~Lett.~\textbf{89}, 130402 (2002).

\bibitem{Milstein02}
  J.N. Milstein, S.J.J.M.F. Kokkelmans, and M.J. Holland, 
  Phys. Rev. A \textbf{66}, 043604 (2002).
  
\bibitem{Perali03}
  A. Perali, P. Pieri, and G.C. Strinati, 
  Phys. Rev. A \textbf{68}, 031601 (2003).

\bibitem{Combescot03}
  R. Combescot, Phys.~Rev.~Lett.~\textbf{91}, 120401 (2003).

\bibitem{Viverit04-1}
  L. Viverit, S. Giorgini, L.P. Pitaevskii, and S. Stringari,
  Phys. Rev. A \textbf{69}, 013607 (2004).
  
\bibitem{Bruun04-1}
  G.M. Bruun and C.J. Pethick, 
  Phys.~Rev.~Lett.~\textbf{92}, 140404 (2004).

\bibitem{Carr04}
  L.D. Carr, G.V. Shlyapnikov, and Y. Castin, 
  Phys.~Rev.~Lett.~\textbf{92}, 150404 (2004).

\bibitem{Karpiuk04}
  T. Karpiuk, M. Brewczyk, and K. Rz{\c a}\.zewski, Phys. Rev. A
  \textbf{69}, 043603 (2004).

\bibitem{Perali04-1}
  A. Perali, P. Pieri, L. Pisani, and G.C. Strinati, 
  Phys. Rev. Lett.~\textbf{92}, 220404 (2004).

\bibitem{Stajic04}
  J. Stajic, J.N. Milstein, Q. Chen, M.L. Chiofalo, M.J. Holland, and
  K. Levin, Phys. Rev. A \textbf{69}, 063610 (2004).

\bibitem{DePalo04}
  S. De Palo, M.L. Chiofalo, M.J. Holland, and S.J.J.M.F. Kokkelmans,
  Phys. Lett. A \textbf{327}, 490 (2004).

\bibitem{Heiselberg04}
  H. Heiselberg, Phys.~Rev.~Lett.~\textbf{93}, 040402 (2004).

\bibitem{Buchler04}
  H.P. B\"uchler, P. Zoller, and W. Zwerger, 
  Phys.~Rev.~Lett.~\textbf{93}, 080401 (2004).

  %\bibitem{Fuchs04}
  %J.N. Fuchs, A. Recati, and W. Zwerger, Phys.~Rev.~Lett.~\textbf{93},
  %090408 (2004).

\bibitem{Perali04-2}
  A. Perali, P. Pieri, and G.C. Strinati,
  Phys. Rev. Lett.~\textbf{93}, 100404 (2004).

\bibitem{Kinnunen04}
  J. Kinnunen, M. Rodriguez, and P. Torma, 
  Science \textbf{305}, 1131 (2004).
  
\bibitem{Williams04}
  J.E. Williams, N. Nygaard, and C.W. Clark, 
  New J. Phys. \textbf{6}, 123 (2004).
  
\bibitem{Drummond04}
  P.D. Drummond and K.V. Kheruntsyan, 
  Phys. Rev. A \textbf{70}, 033609 (2004).

\bibitem{Avdeenkov05}
  A.V. Avdeenkov and J.L. Bohn, Phys. Rev. A \textbf{71}, 023609
  (2005).

\bibitem{Parish05}
M.M. Parish, B. Mihaila, E.M. Timmermans, K.B. Blagoev, and
P.B. Littlewood, Phys. Rev. B \textbf{71}, 064513 (2005).
  
\bibitem{Bulgac04}
  A. Bulgac and G.F. Bertsch, cond-mat/0404301 (2004).
  
\bibitem{Diener04-1}
  R.B. Diener and T.-L. Ho, cond-mat/0404517 (2004).
  
\bibitem{Diener04-2}
  R.B. Diener and T.-L. Ho, cond-mat/0405174 (2004).
  
\bibitem{Jensen04}
  L.M. Jensen, cond-mat/0412431 (2004).
  
  %\bibitem{Viverit04}
  %L. Viverit, G.M. Bruun, A. Minguzzi, and R. Fazio,
  %Phys.~Rev.~Lett.~\textbf{93}, 110406 (2004).
  
\bibitem{Negele98}
  J.W.~Negele and H.~Orland, {\em Quantum many-particle systems}\,
  (Perseus, Cambridge, Massachusetts, 1998).
  
\bibitem{KGKB04}
  K. G\'{o}ral and K. Burnett, Phys. World \textbf{17} (3), 23 (2004).  
  
\end{thebibliography}
\end{document}